\documentclass[showpacs,preprintnumbers,amsfonts,amssymb,amsmath,nofootinbib]{revtex4}
\usepackage{graphicx}
\usepackage{dcolumn}
\usepackage{bm}

\begin{document}
\title{Electron and Muon production cross sections in quasielastic $\nu(\bar\nu)$-Nucleus scattering for $E_\nu < 1~GeV$}
\author{F. Akbar, M. Rafi Alam, M. Sajjad Athar, S. Chauhan, S. K. Singh and F. Zaidi\footnote{Corresponding author: zaidi.physics@gmail.com} }
\affiliation{Department of Physics, Aligarh Muslim University, Aligarh-202 002, India}
\date{\today}
\begin{abstract}
In this work, we have studied (anti)neutrino induced charged current 
quasielastic scattering from some nuclear targets in the energy
 region of $E_\nu < 1~GeV$. Our aim is to confront electron and muon production
 cross sections relevant for $\nu_\mu \leftrightarrow \nu_e$ or $\bar\nu_\mu \leftrightarrow 
\bar\nu_e$ oscillation experiments. The effects due to lepton mass and its kinematic implications, radiative corrections, 
 second class currents and uncertainties in the axial and pseudoscalar form factors are calculated for (anti)neutrino induced reaction cross sections on
 free nucleon as well as the nucleons bound in a nucleus where nuclear medium effects influence the cross section.
 For the nuclear medium effects we have taken some versions of Fermi gas model(FGM) available in literature. The results 
 for (anti)neutrino-nucleus scattering cross section per interacting nucleons are compared with the corresponding results in free nucleon case. 
 \end{abstract}
\maketitle

\section{Introduction}
With the measurement of $\theta_{13}$ at the nuclear 
reactors~\cite{Abe:2011fz,Ahn:2012nd,An:2012eh} and finding a clear evidence for 
$\bar\nu_{e}$ disappearance, now the emphasis is upon determining mass hierarchy in the neutrino 
sector as well as to find signals of CP violation in the leptonic sector~\cite{Ericson:2015cva, King:2013eh}.
 For these physics goals, accelerator experiments like T2K~\cite{Abe:2011ks}, 
NO$\nu$A~\cite{Habig:2012uva}, etc.
 are taking data and looking for $\nu_\mu \leftrightarrow \nu_e$ or $\bar\nu_\mu 
\leftrightarrow \bar\nu_e$ oscillation signals and experiments like LBNO~\cite{Agarwalla:2014ura}, 
 DUNE~\cite{dune}, T2HK~\cite{Ishida:2013kba, Abe:2015zbg}, etc. are planned.
  Most of these oscillation experiments are being 
performed in the (anti)neutrino energy region of $\sim1~GeV$. 
 In this energy region the major contribution to the event rates comes from the 
charged current quasielastic(CCQE) lepton production process followed by 
 charged current induced one pion production process. 
 There are other channels which may contribute to the lepton event rates like
strange particle production through 
$|\Delta S|$=1 processes but they are Cabibbo suppressed, 
 while their production through strangeness conserving $|\Delta S|$=0 processes are suppressed due to the threshold effect.
  The contribution to the lepton event rates from the deep inelastic scattering 
is expected to be small. 
The importance of CCQE lepton production process is two fold. Firstly, the 
observation of charged lepton in the final state is the cleanest signature of a (anti)neutrino 
 interaction and secondly  this is the 
simplest process using which energies of the incoming (anti)neutrinos 
 may be determined. However, there are theoretical uncertainties involved while inferring 
the energy of neutrinos even from CCQE processes on nuclear targets 
 and this has been discussed in detail recently by many authors~\cite{Gran:2013kda,Mosel:2013fxa,
 Benhar:2015wva,Nieves:2011yp,Martini:2011wp,Martini:2012uc,Ankowski:2014yfa,Lalakulich:2012hs}. 
 
 In the electron and muon (anti)neutrinos processes, the contribution to the cross section 
from CCQE process  would be different for the production of electrons and muons, due to lepton mass and other effects 
even in the presence of the universality of weak interaction. 
Recently, Day and 
McFarland~\cite{Day:2012gb} studied the effect of lepton mass, radiative corrections and 
uncertainties in the nucleon electroweak form factors including the second class currents(SCC) 
on the (anti)neutrino CCQE scattering cross sections from the nucleon targets.
 In their work, it has been shown that the radiative corrections at the tree 
level CCQE process may lead to important difference between electron and muon 
 production cross sections~\cite{Day:2012gb}, as it is proportional to 
$\log(\frac{E_l^\ast}{m_l})$, where $E_l^\ast$ is the outgoing lepton energy in the center of mass frame and $m_l$ is the mass of the charged lepton. 
Furthermore, the variation in the axial dipole mass $M_A$ which has been recently 
 discussed in literature will also lead to difference in electron and muon cross sections.
 The variation in the experimental measurements of $M_A$ has been  recently found out to be quite large i.e. $0.99~GeV$ 
to $1.35~GeV$~\cite{Fields:2013zhk,Lyubushkin:2008pe,AguilarArevalo:2010zc,AguilarArevalo:2010cx,
AguilarArevalo:2013hm,Fiorentini:2013ezn,Abe:2013jth,Nakajima:2010fp,Gran:2006jn,Espinal:2007zz},
 from the world average($1.026\pm0.021~GeV$~\cite{Bernard:2001rs} and $1.014\pm0.014~GeV$~\cite{Bodek:2007ym}).
  A change of 10$\%$ in the value of $M_A$ say from $1~GeV$ to $1.1~GeV$ would 
result in almost an increase of 10$\%$ in the CCQE scattering cross section.
Generally, the pseudoscalar form factor $F_P(Q^2)$ is expressed in terms of axial vector form factor 
$F_A(Q^2)$ using the Goldberger-Trieman relation and PCAC. However, other parameterizations of the pseudoscalar form factor $F_P(Q^2)$  based on
Chiral Perturbation Theory and Lattice calculations have also been discussed in literature
~\cite{Bernard:1994wn,Bernard:1998gv, Bernard:2000et,Schindler:2006it,Tsapalis:1900zz}.  Moreover, if 
one considers the presence of second class currents, then there would be additional contribution to the 
(anti)neutrino nucleon cross sections due to the form factors $F_3^V(Q^2),~ F_3^A(Q^2)$ associated with them.
The inclusion or noninclusion of these contributions may translate into the systematic
  uncertainties in the determination of event rates. 
Some of these effects have not been taken into account in the 
 (anti)neutrino Monte Carlo generators like GENIE~\cite{Andreopoulos:2009rq}, 
NEUT~\cite{Hayato:2009zz}, NUANCE~\cite{Casper:2002sd}, NuWro~\cite{Golan:2012rfa}, GiBUU~\cite{Lalakulich:2011eh}, etc.  
 
In almost all the present generation (anti)neutrino experiments, moderate to heavy nuclear targets 
like $^{12}C$, $^{16}O$, $^{40}Ar$, $^{56}Fe$, $^{208}Pb$, etc. are being used. 
Recently, the measurements have been performed for $Q^2$ 
distribution~\cite{Lyubushkin:2008pe, AguilarArevalo:2010zc,AguilarArevalo:2010cx, AguilarArevalo:2013hm, Fields:2013zhk, Fiorentini:2013ezn}, 
double-differential scattering cross sections~\cite{AguilarArevalo:2010zc, AguilarArevalo:2013hm, Abe:2013jth} and 
total scattering cross sections~\cite{Lyubushkin:2008pe, AguilarArevalo:2010zc, AguilarArevalo:2013hm, Nakajima:2010fp, Abe:2013jth} 
using some of these nuclear targets. 
In these nuclear targets nuclear medium effects play 
an important role and the effect has been found to be substantial in the low 
energy region. Most of the present generation Monte Carlo generators are using
relativistic Fermi gas model given by  
Smith and Moniz~\cite{Smith} or the model discussed by Llewellyn Smith~\cite{LlewellynSmith:1971zm}. 
However, there are other variants of Fermi gas 
model available in literature like that of Gaisser and O'Connell~\cite{Gaisser:1986bv}, Singh 
and Oset~\cite{Singh:1992dc,Singh:1993rg}, Nieves et al.~\cite{Nieves:2004wx}, etc. which have been applied to study (anti)neutrino-nucleus reactions. 
In addition to these works many other theoretical models like superscaling approach\cite{Megias:2014qva}, mean field approximation~\cite{Amaro:2006pr}, 
relativistic meson-nucleon model~\cite{Kim:1994zea}, relativistic Green's function model~\cite{Meucci:2014bva}, plane wave impulse 
approximation(PWIA)~\cite{Benhar:2005dj, Ankowski:2005wi}, distorted wave impulse 
approximation(DWIA)~\cite{Butkevich:2012zr}, etc., have been used in literature. Moreover, 
in the context of Fermi gas models the effect of 
two particle-two hole(2p-2h) correlation, meson exchange currents and multinucleon
mechanism have also been considered in literature~\cite{Martini:2009uj,Nieves:2011pp,Simo:2014wka,Lovato:2014eva}. These studies  have been recently summarized in some review 
articles~\cite{Benhar:2015wva,Morfin:2012kn,Alvarez-Ruso:2014bla}, but they have not been incorporated into the present
Monte Carlo generators being used in 
the neutrino oscillation experiments~\cite{Mosel:2015yaa}.  
  
In the present work, we have studied $\nu_l,~\bar\nu_l;~(l=e,~\mu)$ induced CCQE 
scattering from some nuclear targets like $^{12}C$, $^{40}Ar$, $^{56}Fe$ and 
$^{208}Pb$ in the energy region of $E_\nu<1~GeV$ including the effect due to lepton mass and its kinematic implications, 
radiative corrections, form factors, second class 
currents, etc. We have performed the calculations using 
Local Fermi Gas Model(LFG)~\cite{SajjadAthar:2009rd,SajjadAthar:2009rc,Athar:2005hu,Athar:1999fu} 
 and also compared the numerical results with the different Fermi gas models of 
Smith and Moniz~\cite{Smith}, Llewellyn Smith~\cite{LlewellynSmith:1971zm}, 
Gaisser and O'Connell~\cite{Gaisser:1986bv}. 
Using the different nuclear models we have studied the difference in the 
lepton(electron vs muon) cross sections due to the axial dipole mass, pseudoscalar form factor,
radiative corrections and effect of second class currents. 
Furthermore, in the Local Fermi Gas Model~\cite{SajjadAthar:2009rd,SajjadAthar:2009rc,Athar:2005hu,Athar:1999fu} we have also 
included the nucleon-nucleon interactions due to which response 
of electroweak transition strength is modified. These modifications are calculated by 
incorporating the interactions of particle-hole(1p-1h) excitation 
in the nuclear medium in a random phase 
approximation(RPA)~\cite{Singh:1992dc,Nieves:2004wx}. 
Our aim of this work is to study the role of dominant nuclear medium effects like Fermi motion,
binding energy and nucleon correlations on the various physics inputs which lead to difference 
in electron and muon scattering cross sections in the case of per nucleon target as discussed by Day and McFarland~\cite{Day:2012gb}.
In view of this, the effect of 2p-2h, meson exchange currents and multinucleon mechanism
have not been considered. This 
is at present beyond the scope of this paper and it may 
be studied as a separate work. In section-\ref{Formalism}, we present the formalism in brief, in 
section-\ref{Results and Discussion}, results and discussions are presented and
the findings are summarized in section-\ref{sec:summary}.

\section{Formalism}\label{Formalism}
The basic reaction for the quasielastic process is a (anti)neutrino interacting with a (proton)neutron target given by
\begin{eqnarray}\label{quasi_reaction}\left.
\begin{array}{l}
\nu_l(k)~+~n(p)~\rightarrow~l^-(k^{\prime})~+~p(p^\prime) \\
{\bar{\nu}}_l(k)~+~p(p)~\rightarrow~l^+(k^{\prime})~+~n(p^\prime) 
\end{array}\right\}~~l=e,\mu
\end{eqnarray}
where $k,~k^\prime$ are the four momenta of incoming and outgoing lepton and $p,~p^\prime$ are the four momenta of initial and final
nucleon, respectively.
The invariant matrix element for the charged current reaction of (anti)neutrino, given by Eq.(\ref{quasi_reaction}) is written as
\begin{eqnarray}\label{qe_lep_matrix}
{\cal M}=\frac{G_F}{\sqrt{2}}\cos\theta_c~l_\mu~J^\mu
\end{eqnarray}
where $G_F$ is the Fermi coupling constant (=1.16639$\times 10^{-5}~GeV^{-2}$), 
$\theta_c(=13.1^0)$ is the Cabibbo angle. The leptonic weak current is 
given by
\begin{eqnarray}\label{lep_curr}
l_\mu&=&\bar{u}(k^\prime)\gamma_\mu(1 \pm \gamma_5)u(k),
\end{eqnarray}
where ($+$ve)$-$ve sign is for (antineutrino)neutrino.
$J^\mu$ is the hadronic current given by
\begin{eqnarray}\label{had_curr}
J^\mu&=&\bar{u}(p')\Gamma^\mu u(p),
\end{eqnarray}
with
\begin{eqnarray}\label{eq:had_int}
\Gamma^\mu&=&F_1^V(Q^2)\gamma^\mu+F_2^V(Q^2)i\sigma^{\mu\nu}\frac{q_\nu}{2M} 
+ F_3^V(Q^2)\frac{q^\mu}{M} \nonumber \\
&+& F_A(Q^2)\gamma^\mu\gamma^5 
+ F_P(Q^2) \frac{q^\mu}{M}\gamma^5  + F_3^A(Q^2)\frac{(p+p^\prime)^\mu}{M}\gamma^5,
\end{eqnarray}
 $Q^2(=-q^2)~\geq 0$ is the four momentum transfer square and $M$ is the nucleon mass. $F_{1,2}^V(Q^2)$
 are the isovector vector form factors and
 $F_A(Q^2)$, $F_P(Q^2)$ are the axial and pseudoscalar form factors, respectively. 
$F_3^V(Q^2)$ and $F_3^A(Q^2)$ are the form factors
related with second class current.
Using the leptonic and hadronic currents given in Eq.(\ref{lep_curr}) and Eq.(\ref{had_curr}),
the matrix element square is obtained by using
Eq.(\ref{qe_lep_matrix}) as
\begin{equation}\label{mat_quasi}
{|{\cal M}|^2}=\frac{G_F^2}{2}\cos^2\theta_c~{ L}_{\mu\nu} {J}^{\mu\nu}
\end{equation} 
${ L}_{\mu\nu}$ is the leptonic tensor calculated to be
\begin{eqnarray}\label{lep_tens}
{L}_{\mu\nu}&=&{\bar\Sigma}\Sigma{l_\mu}^\dagger l_\nu=L_{\mu\nu}^{S} \pm i L_{\mu\nu}^{A},~~~~\mbox{where}\\
L_{\mu\nu}^{S}&=&8~\left[k_\mu k_\nu^\prime+k_\mu^\prime k_\nu-g_{\mu\nu}~k\cdot k^\prime\right]~~~~\mbox{and}\nonumber\\
L_{\mu\nu}^{A}&=&8~\epsilon_{\mu\nu\alpha\beta}~k^{\prime \alpha} k^{\beta},
\end{eqnarray}
where the $+$ sign($-$ sign)  is for neutrino(antineutrino).

The hadronic tensor ${J}^{\mu\nu}$ is given by:
\begin{eqnarray}\label{had_tens}
J^{\mu\nu}&=&\bar{\Sigma}\Sigma J^{\mu\dagger} J^\nu\nonumber\\
&=&\frac{1}{2}\mbox{Tr}\left[({\not p^\prime}+M)\Gamma^\mu ({\not p}+M)\tilde\Gamma^\nu\right]
\end{eqnarray}
where $\tilde\Gamma^\nu=\gamma^0~{\Gamma^\nu}^\dagger~\gamma^0$.\\
The hadronic current contains isovector vector form factors $F_{1,2}^V(Q^2)$ of the nucleons, which are given as
\begin{equation}\label{f1v_f2v}
F_{1,2}^V(Q^2)=F_{1,2}^p(Q^2)- F_{1,2}^n(Q^2) 
\end{equation}
where $F_{1}^{p(n)}(Q^2)$ and $F_{2}^{p(n)}(Q^2)$ are the Dirac and Pauli form factors of proton(neutron) 
which in turn are expressed in terms of the
experimentally determined Sach's electric $G_E^{p,n}(Q^2)$ and magnetic $G_M^{p,n}(Q^2)$ form factors as 
\begin{eqnarray}\label{f1pn_f2pn}
F_1^{p,n}(Q^2)&=&\left(1+\frac{Q^2}{4M^2}\right)^{-1}~\left[G_E^{p,n}(Q^2)+\frac{Q^2}{4M^2}~G_M^{p,n}(Q^2)\right]\\
F_2^{p,n}(Q^2)&=&\left(1+\frac{Q^2}{4M^2}\right)^{-1}~\left[G_M^{p,n}(Q^2)-G_E^{p,n}(Q^2)\right]
\end{eqnarray}
$G_E^{p,n}(Q^2)$ and $G_M^{p,n}(Q^2)$ are taken 
from different parameterizations~\cite{Galster:1971kv,Bradford:2006yz,Budd:2005tm,Bosted:1994tm,Alberico:2008sz}.

The isovector axial form factor is obtained from the quasielastic neutrino and antineutrino scattering 
as well as from pion electroproduction data and is parameterized as
\begin{equation}\label{fa}
F_A(Q^2)=F_A(0)~\left[1+\frac{Q^2}{M_A^2}\right]^{-2};~~F_A(0)=-1.267.
\end{equation}

The pseudoscalar form factor is determined by using PCAC which gives a relation between $F_P(Q^2)$ 
and pion-nucleon form factor 
$g_{\pi NN}(Q^2)$ and is given by~\cite{LlewellynSmith:1971zm}:
\begin{equation}
 F_P(Q^2)=\frac{2 M^2 F_A(0)}{Q^2}\left(\frac{F_A(Q^2)}{F_A(0)}-\frac{m_\pi^2}{(m_\pi^2+Q^2)}
 \frac{g_{\pi NN}(Q^2)}{g_{\pi NN}(0)} \right),
\end{equation}
where $m_\pi$ is the pion mass and $g_{\pi NN}(0)$ is the pion-nucleon strong coupling constant. 

$F_P(Q^2)$ is dominated by  the pion pole and is given in terms of axial vector form factor $F_A(Q^2)$
using the Goldberger-Treiman(GT) relation~\cite{LlewellynSmith:1971zm} 
\begin{equation}\label{fp}
F_P(Q^2)=\frac{2M^2F_A(Q^2)}{m_\pi^2+Q^2}.
\end{equation}

The form of pseudoscalar form factor $F_P(Q^2)$ using PCAC may also be written as~\cite{Schindler:2006it}
\begin{eqnarray}\label{fp1}
 F_p(Q^2)&=&\frac{M}{Q^2}\left[\left(\frac{2 m_\pi^2 F_\pi}{m_\pi^2+Q^2}\right)~\left(\frac{M F_A(0)}{F_\pi}+\frac{g_{\pi NN}(0)\Delta Q^2}{m_\pi^2} \right)
 +2 M F_A(Q^2) \right],
\end{eqnarray}
where $g_{\pi NN}(0)=13.21$, $F_\pi=92.42~MeV$ and $\Delta=1+\frac{M F_A(0)}{F_\pi g_{\pi NN}(0)}$.

Pseudoscalar form factor using Chiral Perturbation Theory(ChPT) is given by~\cite{Bernard:1994wn, Bernard:2001rs, Schindler:2006it,Tsapalis:1900zz}
\begin{equation}\label{fp2}
F_P(Q^2)=\frac{2M g_{\pi NN}(0) F_\pi}{m_\pi^2+Q^2}+\frac{F_A(0) M^2 r_A^2}{3}
\end{equation}
where axial radius $r_A=\frac{2\sqrt{3}}{M_A}$.

The form factors $F_3^V(Q^2)$ and $F_3^A(Q^2)$ are associated with the second class current(SCC). There are no 
compelling reasons for their existence as they violate charge or time symmetry and in the case of $F_3^V(Q^2)$
 also the conserved vector current. Almost all the current calculations of neutrino reactions assume SCC to be zero. However,
 there are some experimental analyses of semileptonic weak interactions like beta decays, muon capture and
 neutrino scattering in the $|\Delta S|=0$ sector which give upper limits on these form factors which are consistent with 
 the constraints of the present data on these processes~\cite{Day:2012gb, Ahrens:1988rr, Bhattacharya:2011qm, Cirigliano:2013xha}.
  In Ref.\cite{Ahrens:1988rr}, the upper limit for second class vector current obtained from neutrino experiments is $1.9$. 
We have used the following expressions for $F_3^V(Q^2)$ as given in Ref.~\cite{Ahrens:1988rr}
\begin{equation}\label{eq:f3v1}
F_3^V(Q^2)=\frac{F_3^V(0)}{\left(1+\frac{Q^2}{M_3^{V2}}\right)^2}.
\end{equation}
To observe the maximum effect of second class vector current we have taken $F_3^V(0)=1.6$ on the upper side of the limit with $M_3^V=1~ GeV$~\cite{Ahrens:1988rr}
for our numerical calculations.
Another expression for $F_3^V(Q^2)$ as given in Ref.\cite{Day:2012gb} is
\begin{equation}\label{eq:f3v2}
F_3^V(Q^2)=4.4~F_1^V(Q^2).
\end{equation}
The axial form factor associated with the second class
current $F_3^A(Q^2)$  is  taken as~\cite{Day:2012gb, Ahrens:1988rr} 
\begin{equation}\label{eq:f3a1}
F_3^A(Q^2) =0.15~F_A(Q^2). 
\end{equation}
The parameterization of form factors discussed above will be used in the evaluation 
of the CCQE cross section.
The differential scattering cross section for reaction given in Eq.(\ref{quasi_reaction}) in the laboratory frame 
is in general written as,
\begin{eqnarray}\label{diff_xsect_quasi}
d\sigma=\frac{(2\pi)^{4}\delta^{4}(k+p-p^\prime-k^\prime)}{4(k\cdot p)
}\frac {d^{3}{\vec{k^\prime}}}{(2\pi)^{3}2E_l}\frac {d^{3}{\vec {p^\prime}}}{(2\pi)^{3}2E_{p}} {\bar\Sigma}\Sigma| \mathcal{M} |^2.
\end{eqnarray}

The double differential cross section  $\sigma_{free}(E_l,\Omega_l)$ on free nucleon is then obtained as
\begin{equation}\label{sig_zero}
\sigma_{free}(E_l,\Omega_l)\equiv \frac{d^2 \sigma}{ d E_l \; d \Omega_l }=
\frac{{|\vec k^\prime|}}{64\pi^2 E_\nu E_n E_p }{\bar\Sigma}\Sigma{|{\cal M}|^2}\delta[q_0+E_n-E_p]
\end{equation}
Inside the nucleus, the neutrino scatters from a neutron moving in a finite nucleus of neutron density $\rho_n(r)$, with a 
local occupation number $n_n({\vec{p}},{\vec{r}})$  of the initial
nucleon of momentum $\vec p$ localized in a region $r$, the radius of the nucleus.
In the local density approximation the scattering cross section is written as 
\begin{equation}\label{sig_4}
\sigma(E_l,\Omega_l) =
\int 2d{\vec r}d{\vec p}\frac{1}{(2\pi)^3}n_n({\vec p},{\vec r})\sigma_{free}(E_l,\Omega_l)
\end{equation}
where $\sigma_{free}(E_l,\Omega_l)$ is given by Eq.(\ref{sig_zero}). The neutron energy $E_n$ and proton energy $E_p$ are 
replaced by $E_n(|\vec p|)$ and $E_p(|\vec{p}+\vec{q}|)$, where $\vec{p}$ is now the momentum of the target nucleon  inside the nucleus. 
This is because inside the nucleus the nucleons are not free and their momenta are constrained to 
satisfy the Pauli principle, i.e., ${p}<{p_{F_{n}}}$ and ${p^\prime}(=|{\vec p}+{\vec q}|) > p_{F_{p}}$, where
$p_{F_n}$ and $p_{F_p}$ are the local Fermi momenta of neutrons and protons at the interaction point in the nucleus and are 
given by $p_{F_n}=\left[3\pi^2\rho_n(r)\right]^\frac{1}{3}$ and $p_{F_p}=\left[3\pi^2\rho_p(r)\right]^\frac{1}{3}$, $\rho_n(r)$ and 
$\rho_p(r)$ are the neutron and proton nuclear densities which are given in terms of the nuclear density $\rho(r)$ :
\begin{equation}\label{rho}
\rho_{n}(r)=\frac{(A-Z)}{A}\rho(r);~~~\rho_{p}(r)=\frac{Z}{A}\rho(r).
\end{equation}
The density parameters have been taken from
Ref.~\cite{Vries,GarciaRecio:1991wk} and are summarized in Table-\ref{tab:nuc_para}. For the 
antineutrino induced reaction on free nucleon or nucleons bound in a nucleus the role 
of neutron and proton get interchanged. Furthermore, in nuclei
the threshold value of the reaction i.e. the $Q-$value of 
the reaction($Q_{r}$) has to be taken into account, which we have 
taken to be the value corresponding to the lowest allowed Fermi transition or Gammow-Teller transition.
\begin{table}
 \begin{center}
\begin{tabular}{lccccccccc} \\  \hline \hline 
   & \multicolumn{2}{c}{ $c_1$ }  &   \multicolumn{2}{c}{ $c_2$ }  &   \multicolumn{2}{c}{ $Q-$value } & $B.E.$ &  \multicolumn{2}{c}{$p_F$} \\
   &    $c_1^n$ & $c_1^p$ & $c_2^n$ & $c_2^p$  &  $\nu$ & $\bar \nu$   &   & $\nu$ & $\bar \nu$ \\	\hline 			 
$^{12}C$  &  1.692   & 1.692    & 1.082$^{\ast}$   &   1.082$^{\ast}$   &   16.8  &   13.9   &  25    & 221       &  221\\
$^{40}Ar$ &  3.64   & 3.47    &  0.569     &   0.569    &    2.5    &   8.0   & 	30    & 259  & 242 \\
$^{56}Fe$ &  4.05  & 3.971   &  0.5935    &   0.5935   &    6.8   &   4.8    &  36    & 263 &  251\\
$^{208}Pb$ &  6.89   & 6.624    &  0.549   &  0.549    &    2.4  &   5.5   & 	44    & 283 &  245 \\ \hline \hline 
\end{tabular}
\end{center}
\caption{Different parameters used for  numerical calculations for various nuclei.  
$c_1$ and $c_2$ are the density parameters(in Fermi units)  defined for modified harmonic oscillator as  
$\rho(r)=\rho_0 (1 + c_2 (\frac{r}{c_1} )^2) exp(-(\frac{r}{c_1})^2 )$
and for 2-parameter Fermi density as $\rho(r)=\rho_0 /(1+exp(\frac{r-c_1}{c_2}))$. For $^{12}C$
we have used modified harmonic oscillator density($^{\ast}$ $c_2$ is dimensionless) and for $^{40}Ar$,$^{56}Fe$ and $^{208}Pb$   nuclei,
 2-parameter Fermi density have been used, where superscript $n$ and $p$ in density parameters($c_{i}^{n,p}$; $i$=1,2) stand for neutron and proton, respectively.
 The $Q-$value of the reaction, binding energy ($B.E$)
 and  Fermi momentum($p_F$) for different nuclei are given in MeV. }
 \label{tab:nuc_para}
\end{table}

\begin{center}
\begin{figure}
\includegraphics[height=7 cm, width=12 cm]{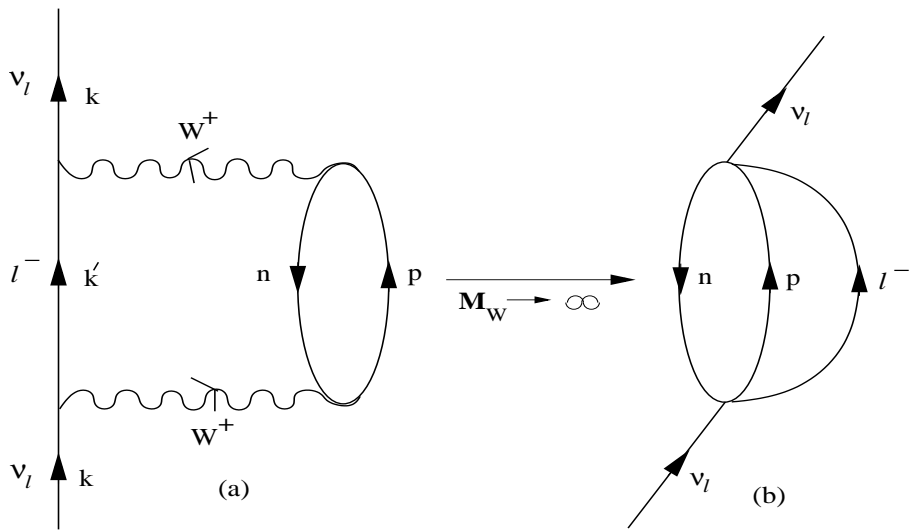}
\caption{Diagrammatic representation of the neutrino self-energy corresponding to 
the ph-excitation leading to $\nu_l +n \rightarrow l^- + p$ in
nuclei. In the large mass limit of the intermediate vector boson(i.e. $M_W\rightarrow \infty$) 
the diagram (a) is reduced to (b) which is used to 
calculate ${|{\cal M}|^2}$ in Eq.(\ref{mat_quasi}).}
\label{fg:fig1}
\end{figure}
\end{center}

These considerations lead to a modification in the $\delta$ function used in Eq.(\ref{sig_zero}) i.e.  $\delta[q_0+E_n-E_p]$ is 
modified to $\delta[q_0+E_n(\vec{p})-E_p(\vec{p}+\vec{q})-Q_{r}]$ and the factor
\begin{equation}\label{delta_modi}
\int \frac{d\vec{p}}{(2\pi)^3}{n_n(\vec{p},\vec{r})}\frac{M^2}{E_n E_p}\delta[q_0+E_n-E_p]
\end{equation}
occurring in Eq.(\ref{sig_4}) is replaced by $-(1/{\pi})$Im${{U_N}(q_0,\vec{q})}$, where ${{U_N}(q_0,\vec{q})}$ is the 
Lindhard function corresponding to the particle hole(ph) excitation~\cite{Singh:1993rg} 
shown in Fig.(\ref{fg:fig1}) and is given by:
\begin{equation}\label{lindhard}
{U_N}(q_0,\vec{q}) = {\int \frac{d\vec{p}}{(2\pi)^3}\frac{M^2}{E_nE_p}\frac{n_n(p)
\left[1-n_p(\vec p + \vec q) \right]}{q_0+{E_n(p)}-{E_p(\vec p+\vec q)}+i\epsilon}}
\end{equation}
where $q_0$=$E_{\nu}-E_l-Q_{r}$. For the antineutrino reaction the suffix n and p will get interchanged.
The imaginary part of the Lindhard function is obtained to be~\cite{Singh:1993rg}: 
\begin{equation}\label{lindhard_imag}
Im{U_N}(q_0, \vec{q}) = -\frac{1}{2\pi}\frac{M^2}{|\vec{q}|}\left[E_{F_1}-A\right]
\end{equation}
with $Q^2\geq0$, $E_{F_2}-q_0<E_{F_1}$ and $\frac{-q_0+|\vec{q}|{\sqrt{1+\frac{4{M^2}}{Q^2}}}}{2}<{E_{F_1}}$, 
where $E_{F_1}=\sqrt{p{_{F_n}}^2+{M}^2}$, $E_{F_2}=\sqrt{{p_{F_p}}^2+{M}^2}$ and \\
$A$ = $Max\left[M,\;\; E_{F_2}-q_0,\;\; \dfrac{-q_0+|\vec{q}| \sqrt{1+\frac{4{M^2}}{Q^2}}}{2}  \right]$.

With inclusion of these nuclear effects, the total cross section $\sigma(E_\nu)$ is written as
\begin{eqnarray}\label{xsection_medeffects}
\sigma(E_\nu)&=&-2{G_F}^2\cos^2{\theta_c}\int^{r_{max}}_{r_{min}} r^2 dr \int^{{k^\prime}_{max}}_{{k^\prime}_{min}}k^\prime dk^\prime 
\int_{Q^{2}_{min}}^{Q^{2}_{max}}dQ^{2}\frac{1}{E_\nu^{2} E_l} \nonumber \\ 
&\times& L_{\mu\nu}J^{\mu\nu} Im{U_N}[E_\nu - E_l - Q_{r}, \vec{q}].
\end{eqnarray}
In the above expression $r_{min}$ and $r_{max}$ are the minimum and maximum limits of nuclear size. 
In principle $r$ should vary from 0 to $\infty$ but in the local density approximation density dies out at around $10~fm$, therefore,
for our numerical calculations we have taken integration limits for nuclear size from 0 to 10 $fm$. $k^\prime_{min}$ and $k^\prime_{max}$ are 
minimum and maximum values of outgoing lepton momenta.
The energy and momentum of the outgoing lepton get modified due to the Coulomb interaction, which is taken into account in an modified effective momentum 
approximation(MEMA)~\cite{Singh:1993rg}.

\begin{figure}
\begin{center}
\includegraphics{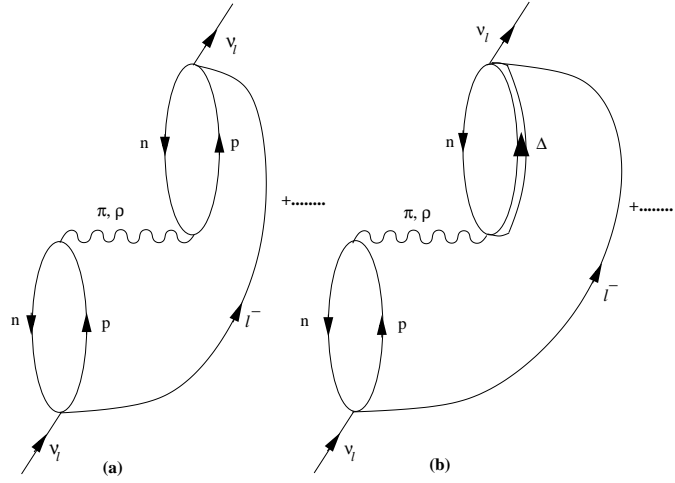}
\caption{Many body Feynman diagrams (drawn in the limit $M_W\rightarrow \infty$) 
accounting for the medium polarization effects contributing to the
process $\nu_l +n \rightarrow l^- + p$.}
\label{fg:fig2}
\end{center}
\end{figure}

In the local density 
approximation, the effective energy of the lepton in the Coulomb field of 
the final nucleus is given by~\cite{Singh:1993rg,Engel:1997fy}:
\[ E_{eff} = E_l + V_c(r), \]
where 
\begin{equation}\label{effective_coulomb}
V_c(r)=4\pi\alpha\;Z_f\left(\frac{1}{r}\int_0^r\frac{\rho_p(r^\prime)}{Z_f}{r^\prime}^2dr^\prime +
\int_r^\infty\frac{\rho_p(r^\prime)}{Z_f}{r^\prime}dr^\prime\right)
\end{equation}
with $\alpha$ as fine structure constant and $Z_f$ as the charge of outgoing lepton, taken as
$-1$  for neutrino and $+1$ for antineutrino. 
This leads to a change in the imaginary part of the Lindhard function occurring in Eq.~(\ref{xsection_medeffects})
\begin{equation*}\label{changed_lindhard}
Im{U_N}[E_\nu - E_l - Q_{r}, \vec{q}] \rightarrow Im{U_N}(E_\nu - E_l - Q_{r} - V_c(r), {\vec q})
\end{equation*}
Furthermore, in a nucleus the response of electroweak strength may change 
due to the presence of strongly
interacting nucleons. These changes are calculated by considering the interaction of ph excitations in the 
nuclear medium in Random Phase Approximation (RPA) as shown in Fig.\ref{fg:fig2}.
The diagram shown in Fig.\ref{fg:fig2} simulates the effects 
of the strongly interacting nuclear medium at the weak vertex. 
The ph-ph interaction is shown by the wavy line in Fig.\ref{fg:fig2} and is 
described by the $\pi$ and $\rho$ exchanges modulated by the effect of short range correlations. 
The effect of the $\Delta$ degrees of freedom in the 
nuclear medium is included in the calculation of the RPA response by considering the effect of
ph-$\Delta$h and $\Delta$h-$\Delta$h excitations as shown 
in Fig.\ref{fg:fig2}(b). This is done by replacing $U_N$ by $U_N=U_N+U_\Delta$, where $U_\Delta$ 
is the Lindhard function for $\Delta$h excitation in the nuclear 
medium and the expressions for $U_N$ and $U_\Delta$ 
are taken from Ref.~\cite{Oset1}. The different couplings of $N$ and $\Delta$ are incorporated in 
$U_N$ and $U_\Delta$. The details of which are
discussed in Refs.~\cite{Singh:1992dc,Nieves:2004wx}. In the Appendix we have given the expression for the
hadronic tensor in covariant form as well as the 
expression for the hadronic tensor when RPA corrections are incorporated. We must point out that the RPA corrections are implemented in the leading order terms only.

Thus, in a local density approximation in the presence of nuclear medium effects including the RPA effect,
the total cross section $\sigma(E_\nu)$, is written as
 \begin{eqnarray}\label{cross_section_quasi}
\sigma(E_\nu)&=&-2{G_F}^2\cos^2{\theta_c}\int^{r_{max}}_{r_{min}} r^2 dr 
\int^{{k^\prime}_{max}}_{{k^\prime}_{min}}k^\prime dk^\prime 
\int_{Q_{min}^{2}}^{Q_{max}^{2}}dQ^2\frac{1}{E_{\nu}^2 E_l}L_{\mu\nu}{J^{\mu\nu}_{RPA}} \nonumber \\
&\times& Im{U_N}[E_{\nu} - E_l - Q_{r} - V_c(r), \vec{q}]
\end{eqnarray}
where $J^{\mu\nu}_{RPA}$ is the modified hadronic tensor when RPA effect is incorporated and the energy transferred to the 
hadronic tensor also gets modified from $q_0=E_\nu-E_l$ to $q_0=E_\nu-E_l-Q_r-V_c$.
Explicit expression of $J^{\mu\nu}_{RPA}$
is given in the Appendix.

Now we discuss in brief the form of the various other Fermi gas model used in literature~\cite{LlewellynSmith:1971zm,Gaisser:1986bv}. 
In the Llewellyn Smith Fermi gas model~\cite{LlewellynSmith:1971zm}, the cross section per nucleon in a nucleus 
is equal to the cross section for a free nucleon i.e. $\sigma_{free}$ defined in Eq.\ref{sig_zero}, 
multiplied by $\left[1-\frac{D}{N}\right]$, where 
\begin{eqnarray}\label{llewellyn_factor}
D&=&Z ~~for~~ 2x < u-v\nonumber\\
&=&\frac{1}{2}A\left\{1-\frac{3x}{4}(u^{2}+v^{2})+\frac{x^3}{2}-
\frac{3}{32x}(u^{2}-v^{2})^{2}\right\}~~~~~~~~for~~ u-v < x < u+v\nonumber\\
&=& 0 ~~~ for ~~x > u+v
\end{eqnarray}
with $x=\frac{|\vec q|}{2 p_{F}}$, $u=(\frac{2N}{A})^{1/3}$, $v=(\frac{2Z}{A})^{1/3}$ 
and $N(=A-Z),~ Z,~ A$ are neutron, proton and mass numbers of the initial
nucleus, respectively. $p_{F}$ is the Fermi momentum and the three momentum transfer $|\vec q|=\sqrt{q_0^2+Q^2}$. 

Smith and Moniz~\cite{Smith} used the following expression for the double differential cross section in the Fermi gas model:
\begin{eqnarray}\label{diff_xsect_smithmoniz}
\frac{d^{2}\sigma}{dk^ \prime d\Omega_l} &=&\frac{G_F^{2} {k^ \prime}^{2} \cos^{2}
(\frac{1}{2}\chi)}{2\pi^{2}M}\left\{W_{2}+[2W_{1}+\frac{m_{l}^{2}}{M^{2}}W_{\alpha}] \tan^{2}(\frac{1}{2}\chi)\right.\nonumber\\
&+&\left.(W_{\beta}+W_{8})m_{l}^{2}/(ME_l\; \cos^{2}(\frac{1}{2}\chi))-2W_{8}/M\; \tan(\frac{1}{2}\chi)\right.\nonumber\\
&\times&\left. \sec(\frac{1}{2}\chi)[-Q^2 \; 
\cos^{2}(\frac{1}{2}\chi)+|{\vec q}|^{2}\; \sin^{2}(\frac{1}{2}\chi)+m_{l}^{2}]^{\frac{1}{2}}\right\},
\end{eqnarray}
where $\cos\chi = \frac{k^\prime}{E_l}  \cos\theta$. The form of $W_i$'s and other  details are given in Ref.~\cite{Smith}.  

Gaisser and O'Connell~\cite{Gaisser:1986bv} have used relativistic response function $R(q,q_0)$, 
in a Fermi gas model to take into account nuclear medium effects, 
the expression for the double differential scattering cross section is given by
\begin{eqnarray}
 \frac{d^2\sigma}{d\Omega_ldE_l}&=&C~\frac{d\sigma_{free}}{d\Omega_l}~R(q,q_0),\nonumber\\
 R(q,q_0)&=&\frac{1}{\frac{4}{3}\pi p_{F_{N}}^3} \int \frac{d^3p_N~M^2}{E_NE_{N^\prime}}
 \delta(E_N + q_0 - E_B - E_{N^\prime}) \theta(p_{F_{N}} - |{\vec p_N}|)\nonumber\\
 &\times&\theta(|{\vec p_N}+{\vec q}| - p_{F_{N^\prime}}),~~~~~~
\end{eqnarray}
where $p_{F_{N}}$ is the Fermi momentum for the initial nucleon, $N,N^\prime$=n or p and $C=A-Z$ for
neutrino induced process and $C=Z$ for the antineutrino 
induced process. $\frac{d \sigma_{free}}{d \Omega_l}$ is the differential scattering cross section for the (anti)neutrino reaction on
free (proton)neutron target and we have used the same
expression for the form factors as used in the LFG for the numerical calculations. 
Different parameters associated with nuclear densities, $Q-$value of the 
reaction, binding energy and Fermi momentum used in the numerical calculations
are summarized in Table-\ref{tab:nuc_para}.  

\section{Results and Discussion}\label{Results and Discussion}
\subsection{Nuclear model dependence}\label{sec:nuclear_model}
\begin{figure}
 \includegraphics[height=15cm,width=12.5 cm]{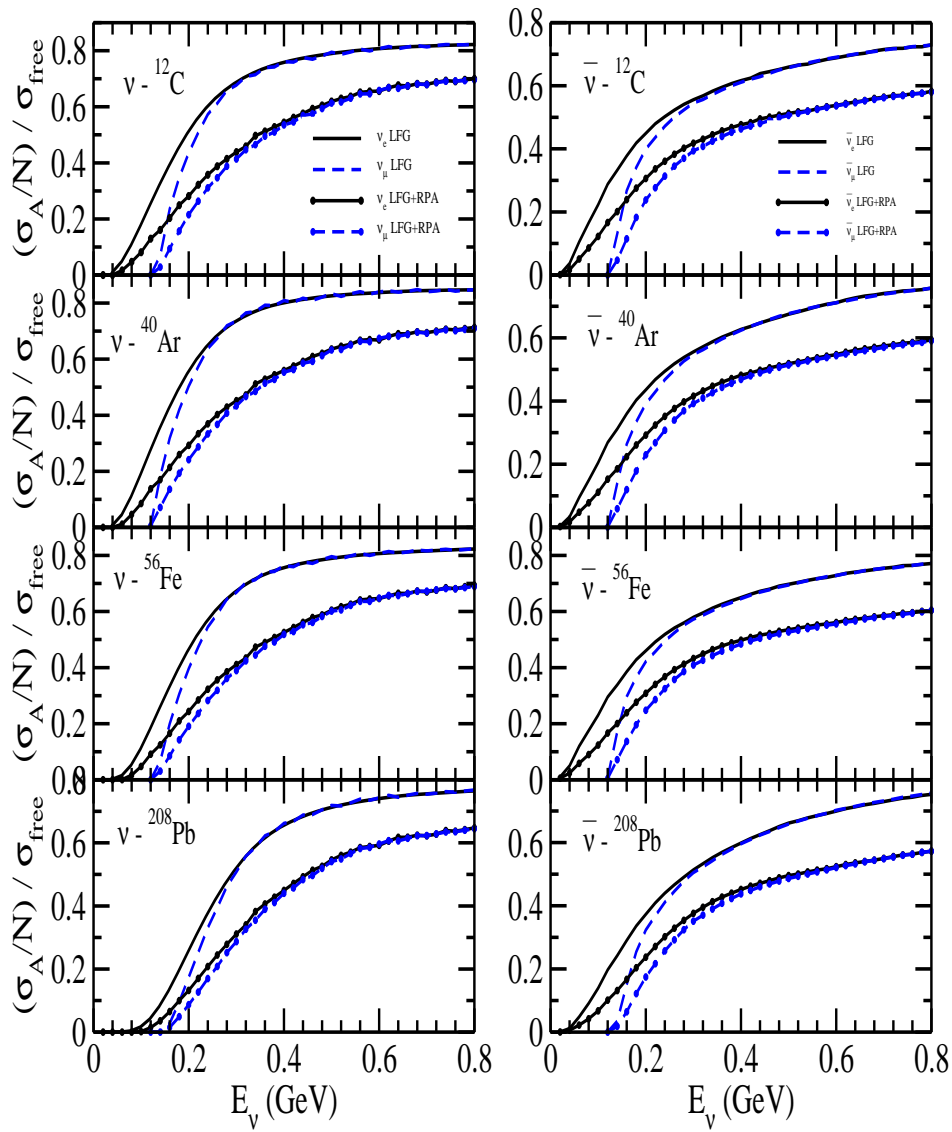}
 \caption{ Ratio $\frac{\sigma_A/N}{\sigma_{free}}$  vs $E_{\nu}$, for neutrino(Left panel) and 
 antineutrino(Right panel) induced processes in $^{12}C$, $^{40}Ar$, $^{56}Fe$ and $^{208}Pb$.
The solid(dashed) line represent cross section obtained from electron(muon) type neutrino and antineutrino beams. 
For neutrino induced process $N=A-Z$, is neutron number 
   and for antineutrino induced process $N=Z$, is proton number. $\sigma_A$ is cross section in nuclear target and has been evaluated using Local 
   Fermi Gas Model(LFG) and LFG with RPA effect(LFG+RPA) 
   and $\sigma_{free}$ is the cross section for the free nucleon case.}
\label{fig3c12}
\end{figure}
In this section, we present the results and discuss the findings. In Fig.~\ref{fig3c12}, the results are presented 
for the ratio of scattering cross section 
per interacting nucleon to the scattering cross section on free nucleon 
target for (anti)neutrino induced processes in $^{12}C$, $^{40}Ar$, $^{56}Fe$ and $^{208}Pb$ in the energy region from threshold to $0.8~GeV$. 
The results are obtained using Local Fermi Gas Model(LFG) i.e. the 
expression given in Eq.\ref{xsection_medeffects} 
and the Local Fermi Gas Model with RPA effect(LFG+RPA) i.e. using Eq.\ref{cross_section_quasi} to the cross section obtained for the free nucleon case 
 using Eq.\ref{diff_xsect_quasi} 
 on neutron(proton) target induced by neutrino(antineutrino). 
 A similar study of $\nu(\bar \nu)-A$ cross sections in the present model 
 at low energies($<0.5 \; GeV$) have also been performed by Kosmas and Oset~\cite{Kosmas:1996fh} in 
 several nuclear targets including $^{40}Ar$. 
 The results obtained in this paper are in agreement with their work. 
 
 Performing calculations using LFG, we find that in $^{12}C$ the nuclear medium effects like Fermi motion, Pauli blocking, binding energy, 
 result in the reduction of cross section by $\sim 30(42)\%$ at 
 $E_{\nu} = 0.3~GeV$ and around $20(30)\%$ at $E_{\nu} = 0.6~GeV$ from free nucleon case for $\nu_e(\bar\nu_e)$ induced processes. 
   Inclusion of RPA correlation in LFG, reduces the cross section for $\nu_e(\bar\nu_e)$ scattering from free nucleon by 
 $\sim 55(56)\%$ at $E_{\nu} = 0.3~GeV$ and $35(45)\%$ at $E_{\nu} = 0.6~GeV$.
Similar results may be observed for $^{40}Ar$, $^{56}Fe$ and $^{208}Pb$ nuclear targets. 
 In general, the reduction in the cross section increases with the 
increase in mass number. For $\nu_{\mu}$ and $\bar \nu_{\mu}$ induced processes at lower energies the reduction is larger 
and for $E_{\nu} > 0.4~GeV$, the reduction in $\nu_{e}$($\bar \nu_{e}$) and $\nu_{\mu}$($\bar \nu_{\mu}$) 
cross sections is almost the same. This will be discussed 
separately in the next section when we compare electron and muon scattering cross sections.

\begin{figure}
\includegraphics[height=10 cm,width=12.5 cm]{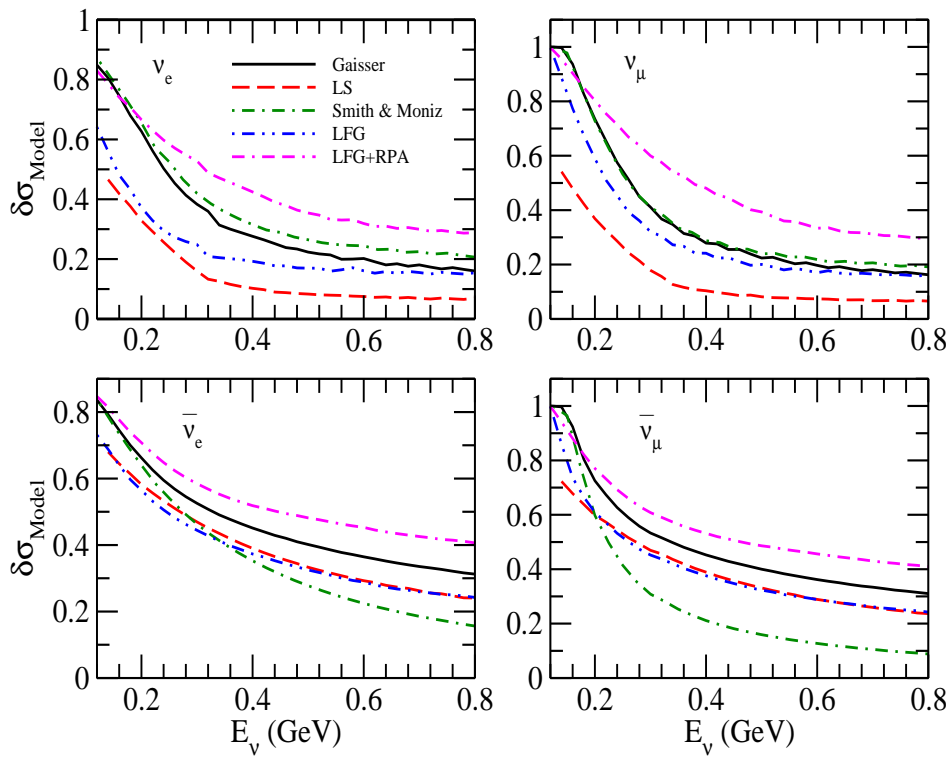}
\caption{The fractional suppression in cross section 
$\delta\sigma_{\rm Model}( = \dfrac{\sigma_{free}-\sigma_{Model}}{\sigma_{free}})$ vs $E_{\nu}$, where
$\sigma_{free}$ is the cross section obtained for free nucleon and $\sigma_{Model}$ is per interacting nucleon cross section in $^{40}Ar$ 
obtained by using different nuclear models.
The results are presented for the cross sections obtained from different 
models of Fermi gas($\sigma_{Model}$) viz. Smith and Moniz~\cite{Smith}(dashed dotted line), Llewellyn Smith~\cite{LlewellynSmith:1971zm}(dashed line), 
Gaisser O' Connell~\cite{Gaisser:1986bv}(solid line), 
and with(double dashed dotted line) \& without RPA(dashed double dotted line) effect using Local Fermi Gas Model. 
The top panel is for neutrino and bottom panel is for antineutrino induced processes.}
\label{fig:del_mod}
\end{figure}
 \begin{figure}
\includegraphics[height=10 cm,width=12.5 cm]{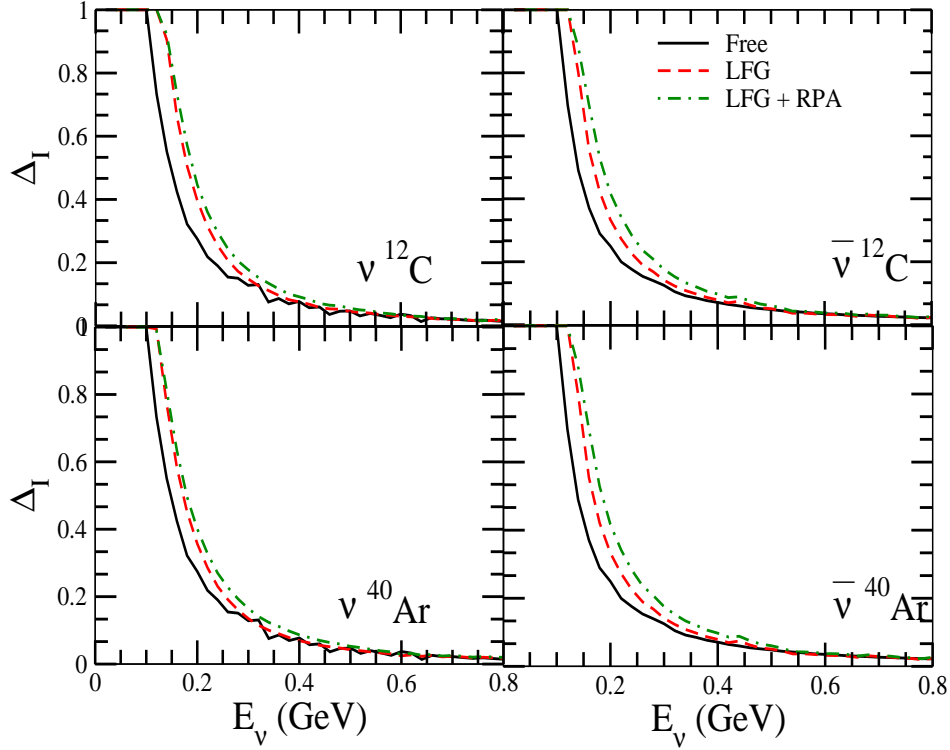}
\caption{$\Delta_{I}= \frac{\sigma_{\nu_e(\bar\nu_e)}-\sigma_{\nu_\mu(\bar\nu_\mu)}}{\sigma_{\nu_e(\bar\nu_e)}} $ for neutrino(left panel)
 and antineutrino(right panel) induced processes in $^{12}C$  and $^{40}Ar$ targets. 
Here $I$ stands for the results of the cross sections obtained (i) for the free nucleon case(solid line)~ (ii) 
in the Local Fermi Gas Model(dashed line) and (iii) for LFG with RPA effect(dashed dotted line).}
\label{fig:delta_sigma}
\end{figure}

To compare our results with other variants of Fermi gas model, 
we have obtained total scattering cross 
section in $^{40}Ar$ using Fermi gas model of Smith and Moniz~\cite{Smith}, Llewellyn 
Smith~\cite{LlewellynSmith:1971zm} and Gaisser and O'Connell~\cite{Gaisser:1986bv} and calculated
 fractional  difference $\delta{\sigma_{Model}}(=\frac{\sigma_{free} - 
\sigma_{Model}}{\sigma_{free}})$, the results for which are shown in Fig.\ref{fig:del_mod}. Here $\sigma_{free}$ stands for the (anti)neutrino induced 
interaction cross section on free nucleon target and $\sigma_{Model}$  stands for the (anti)neutrino induced 
interaction cross section for the nucleons bound inside the nucleus. 
 The results for neutrino is different from 
antineutrino and is mainly due to the interference
 terms with $F_A$ which come with an opposite sign. In the case of LFG with RPA effects, the effect 
of renormalization is large and this suppresses the terms with $F_2$ and $F_A$ which results in a large
change in neutrino vs antineutrino results.
 We find appreciable difference in the results when various nuclear models are 
used. 

 For example, when calculations are performed by using Fermi gas model of Llewellyn Smith~\cite{LlewellynSmith:1971zm}, the cross sections get reduced
 from the free nucleon case by $\sim 16(45)\%$ at $E_{\nu} = 0.3~GeV$ and around $8(30)\%$ 
 at $E_{\nu} = 0.6~GeV$ for $\nu_e$($\bar\nu_e$) induced scattering processes.  
 While when one uses Fermi gas model of Smith and Moniz~\cite{Smith} this reduction in the cross section from the free nucleon case is  
 $\sim 42(45)\%$ at $E_{\nu} = 0.3~GeV$ and $\sim 24(22)\%$ at $E_{\nu} = 0.6~GeV$.
 When the calculations are performed using Fermi gas model of Gaisser and O' Connell~\cite{Gaisser:1986bv} 
 the reduction from the free nucleon case is $\sim 38(52)\%$ at 
 $E_{\nu} = 0.3~GeV$ which becomes $20(35)\%$ at $E_{\nu} = 0.6~GeV$ for $\nu_e(\bar\nu_e)$ induced processes. 
  Performing calculations in the LFG, reduces the cross sections by $\sim 25(44)\%$ at 
 $E_{\nu} = 0.3~GeV$ and around $15(30)\%$ at $E_{\nu} = 0.6~GeV$ for $\nu_e(\bar\nu_e)$ induced processes. 
  Including RPA correlation with LFG, reduces the cross section by 
 $\sim 54(58)\%$ at $E_{\nu} = 0.3~GeV$ and $32(45)\%$ at $E_{\nu} = 0.6~GeV$.
  The nuclear model dependence is found to be larger in the case of $\nu_\mu(\bar\nu_\mu)$ scattering than in the case of 
$\nu_e(\bar\nu_e)$ scattering in the energy region of $E_{\nu}<0.8~GeV$.
 
  For $\nu_\mu(\bar \nu_\mu)$ induced scattering processes, reduction in the cross section from the free nucleon case is  $\sim 18(47)\%$ at 
 $E_{\nu} = 0.3~GeV$ and $\sim 7(30)\%$ at $E_{\nu} = 0.6~GeV$, when cross sections are obtained using Fermi gas model of
 Llewellyn Smith~\cite{LlewellynSmith:1971zm}.   
 Fermi gas model of Smith and Moniz~\cite{Smith} reduces the cross section from the free nucleon case by 
 $\sim 40(30)\%$ at $E_{\nu} = 0.3~GeV$ and around $22(12)\%$ at $E_{\nu} = 0.6~GeV$. 
 When calculations are performed by using Fermi gas model of Gaisser and O' Connell~\cite{Gaisser:1986bv} 
 the reduction from the free nucleon case is $\sim 40(52)\%$ at 
 $E_{\nu} = 0.3~GeV$ which becomes $20(35)\%$ at $E_{\nu} = 0.6~GeV$. 
The calculations are also performed in the LFG and we find that the cross sections get reduced by $\sim 26(45)\%$ at 
 $E_{\nu} = 0.3~GeV$ and around $15(30)\%$ at $E_{\nu} = 0.6~GeV$ for $\nu_\mu(\bar \nu_\mu)$ induced processes. 
  Inclusion of RPA correlation with LFG, reduces the cross section for $\nu_\mu$ scattering from free nucleon value by 
 $\sim 58 \%$ at 
 $E_{\nu} = 0.3~GeV$ and $32 \%$ at $E_{\nu} = 0.6~GeV$, which for $\bar \nu_\mu$ induced scattering is $\sim 60 \%$ at 
 $E_{\nu} = 0.3~GeV$ and $\sim 45 \%$ at $E_{\nu} = 0.6~GeV$.

\subsection{Effect of lepton mass and its kinematic implications}\label{sec:effect_lepmass}
There are two types of corrections which appear when lepton mass $m_l(l=e,\mu)$ is taken into account in the 
cross section calculations for the reaction $\nu_l(\bar\nu_l) + N \rightarrow l^-(l^+) + N^\prime$, ($N,N^\prime=n,p$)
which can be classified as kinematical and dynamical in origin. The kinematical effects arise due to $E_l \ne |\vec k^{\prime}|$ in presence of $m_l$
and the minimum and maximum values of four momentum transfer square ($Q^2=-q^2 \ge 0$) i.e. $Q^2_{min}$ and $Q^2_{max}$ gets modified, affecting the 
calculations of total cross sections. 
These effects are negligible for highly relativistic leptons but could become important at low energies near threshold specially for muons. 
On the other hand, the dynamical corrections arise as additional terms proportional to $\frac{m_l^2}{M^2}$
in the existing contribution of vector and axial vector form factors  as well as new contributions due to induced pseudoscalar and other form factors
associated with the second class currents come into play. In fact all the contributions from the pseudoscalar form factor $F_P(Q^2)$
and the second class vector form factor $F_3^V(Q^2)$ are proportional to $\frac{m_l^2}{M^2}$ while the contribution from the second class axial vector form factor 
$F_3^A(Q^2)$ is proportional either to $\frac{m_l^2}{M^2}$ or $\frac{Q^2}{M^2}$ or both.

To study the lepton mass dependence on  $\nu_e(\bar\nu_e)$ and $\nu_\mu(\bar\nu_\mu)$ 
induced scattering cross sections in free nucleon as well as in nuclear targets, we define
$\Delta_{I}= \frac{\sigma_{\nu_e(\bar\nu_e)}-\sigma_{\nu_\mu(\bar\nu_\mu)}}{\sigma_{\nu_e(\bar\nu_e)}} $
for neutrino/antineutrino induced reaction in $^{12}C$ and $^{40}Ar$ nuclear targets, 
where $I =  i, ~ ii, ~ iii$, which respectively stands for the cross sections obtained in
 ($i$)~ free neutrino/antineutrino-nucleon case, ($ii$)~ the Local Fermi Gas Model(LFG) and ($iii$)~
  the Local Fermi Gas Model with RPA effect(LFG+RPA).  
  
  The results are presented in Fig.\ref{fig:delta_sigma}, which show that the differences in the electron and muon production
  cross sections for $\nu_l(\bar\nu_l)$ induced reactions on $^{12}C$ target are appreciable at 
  low energies $E_{\nu}<0.4~GeV$. For example, 
  this fractional change is about $27(25)\%$ at $E_{\nu}=0.2~GeV$ and reduces to $\sim 8(7)\%$ at $E_{\nu}=0.4~GeV$ in
  the case of free nucleon. While in $^{12}C$, using LFG it is approximately $40(33)\%$ at $E_{\nu}=0.2~GeV$  and  $\sim 8\%$ for 
  both neutrino and antineutrino
  at $0.4~GeV$, respectively.
  However, using RPA effect with LFG, the difference is around 
  $44(42)\%$ and $\sim 9(10)\%$ at $E_{\nu}=0.2~GeV$  and $0.4~GeV$, respectively.

  While for the case of neutrino(antineutrino) induced process on $^{40}Ar$ target using LFG, this fractional change is approximately $35(33)\%$ and
  $\sim 7(8)\%$ at $E_{\nu}=0.2~GeV$  and $0.4~GeV$, respectively. However, using RPA effect with LFG, the difference is around 
  $40(42)\%$ and $\sim 9(10)\%$ at $E_{\nu}=0.2~GeV$  and $0.4~GeV$, respectively. 
\subsection{Form factor dependence}
The hadronic current defined in Eq.\ref{eq:had_int} consists of six form factors;
three isovector($F_i^V(Q^2), i=1,3$) and three axial vector($F_i^A(Q^2), i=1,3$) form factors. 
Among them, $F_i^V(Q^2), i=1,2$, are parameterized in terms of 
Sach's form factors $G_{E}^{p,n}(Q^2)$ and $G_{M}^{p,n}(Q^2)$, 
for which various parameterizations are available in 
literature~\cite{Galster:1971kv, Bradford:2006yz, Budd:2005tm, Bosted:1994tm, Alberico:2008sz}.
$F_3^V(Q^2)$, which arises due to the second class current is generally ignored in calculations. 
Similarly, the axial current consists of three form factors viz. $F_1^A(Q^2)=F_{A}(Q^2)$, $F_2^A(Q^2)=F_{P}(Q^2)$ and $F^{A}_3(Q^2)$,
among them $F_{A}(Q^2)$ is dominant  and is parameterized in a  dipole form with  axial dipole mass($M_A$). 
 The pseudoscalar form factor $F_P(Q^2)$  is given in terms of $F_A(Q^2)$ and $F_{3}^{A}(Q^2)$ arises due to second class currents and is generally ignored.
In a standard calculation of (anti)neutrino nucleon scattering cross sections, form factors associated with first class current are
$F_{1}^{V}(Q^2)$, $F_{2}^{V}(Q^2)$, $F_A(Q^2)$ and $F_P(Q^2)$ and their parameterizations  
are discussed in section~\ref{Formalism}.
We have used these form factors to calculate cross sections and presented the results in section~\ref{sec:nuclear_model} and \ref{sec:effect_lepmass}. However
there are alternate parameterizations of the vector form factors~\cite{Galster:1971kv, Bradford:2006yz, Budd:2005tm, Bosted:1994tm, Alberico:2008sz}
and pseudoscalar form factor and a range of parameter values for the axial dipole mass $M_A$ which have 
also been used in literature to evaluate these cross sections. In the following subsection, we give an estimate of
the uncertainty in the cross sections associated with the use of alternate parameterizations and/or parameter values of these form factors.
 \begin{figure}
  \includegraphics[height=7 cm, width=12 cm]{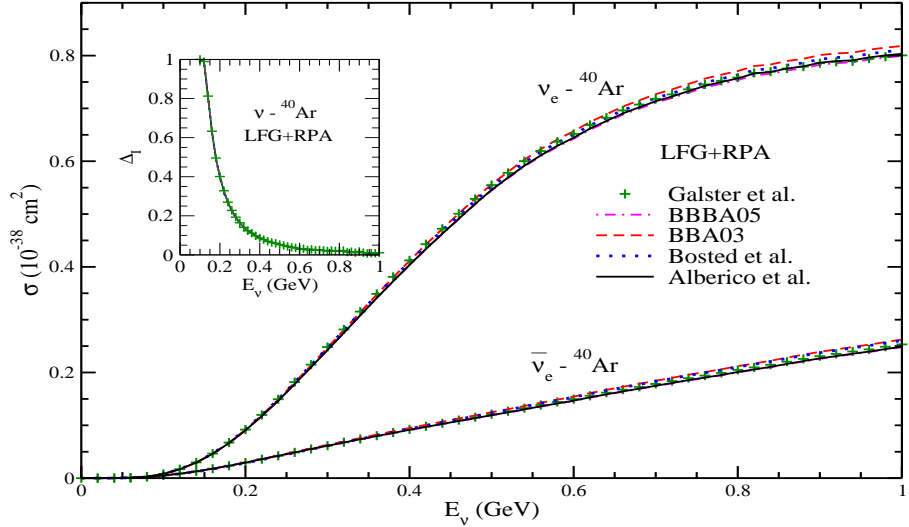}
   \caption{Results of total scattering cross section for $\nu_e(\bar\nu_e)$ induced processes on $^{40}Ar$ are shown using
  different parameterization of the vector form factors~ \cite{Alberico:2008sz, Budd:2005tm, Bradford:2006yz, Bosted:1994tm, Galster:1971kv}.
Symbol $+$ (Galster et al.\cite{Galster:1971kv}), dashed dotted line(Bradford et al.~\cite{Bradford:2006yz}),
dashed line(Budd et al.~\cite{Budd:2005tm}), dotted line(Bosted et al.\cite{Bosted:1994tm}) and 
solid line(Alberico et al.~\cite{Alberico:2008sz})
are the results obtained using various parameterizations for total scattering cross section.
  Inside the inset $\Delta_{I}= \frac{\sigma_{\nu_e}-\sigma_{\nu_\mu}}{\sigma_{\nu_e}}$ for neutrino induced process using these form factors have also been shown.}
  \label{ff_dependence}
 \end{figure}
 \subsubsection{Vector form factors}
 Vector form factors $F_i^V(Q^2), i=1,2$ in Eq.~\ref{f1v_f2v}, are defined in terms of Dirac and Pauli form factors of proton and neutron which in turn are parameterized in terms of 
 experimentally determined Sach's form factors $G_{E}^{p,n}(Q^2)$ and $G_{M}^{p,n}(Q^2)$. For the Sach's form factors  
several parameterizations are discussed in literature by various groups like Bradford et al.~\cite{Bradford:2006yz}, Budd et al.~\cite{Budd:2005tm},
Bosted et al.~\cite{Bosted:1994tm}, Alberico et al.~\cite{Alberico:2008sz}, etc. 
 We have studied
 these parameterizations to observe their effect on scattering cross sections for (anti)neutrino induced CCQE processes on 
nuclear target like $^{40}Ar$
in local density approximation with RPA effect, i.e. LFG+RPA
 and the results are shown in Fig.\ref{ff_dependence}. From the figure it may be observed that
the cross sections obtained using parameterizations of Galster et al.~\cite{Galster:1971kv}, 
 Budd et al.~\cite{Budd:2005tm}, Bosted et al.~\cite{Bosted:1994tm} and Alberico et al.~\cite{Alberico:2008sz} 
 are in agreement with each other 
within $1\%$ which also agrees with the 
recent parameterization discussed by Bradford et al.~\cite{Bradford:2006yz} in the energy region of $0.4 ~GeV-0.8~GeV$.
We have also shown the fractional change
$\Delta_{I}= \frac{\sigma_{\nu_e }-\sigma_{\nu_\mu }}{\sigma_{\nu_e }}$
(in the inset of Fig.\ref{ff_dependence}) for $\nu_l$ induced CCQE process using different parameterizations and found that 
the dependence on the choice of parameterizations for $F_{1}^V(Q^2)$ and $F_{2}^V(Q^2)$ is almost negligible.
  \subsubsection{Axial vector form factor}
  \begin{table}
 \begin{center}
\begin{tabular}{|c|c|c|c|}  \hline 
  Experiment & $M_A~(GeV)$ & Experiment & $M_A~(GeV)$\\ \hline \hline 
   MINER$\nu$A~\cite{Fields:2013zhk, Fiorentini:2013ezn} & 0.99 & SciBooNE~\cite{Nakajima:2010fp} & 1.21$\pm$0.22 \\ \hline
      NOMAD~\cite{Lyubushkin:2008pe} & 1.05$\pm$0.02$\pm$0.06 & K2K-SciBar~\cite{Gran:2006jn} & 1.144$\pm$0.077 \\ \hline
   MiniBooNE~\cite{AguilarArevalo:2010cx, AguilarArevalo:2010zc, AguilarArevalo:2013hm} & 1.23$\pm$0.20 & K2K-SciFi~\cite{Gran:2006jn} & 1.20$\pm$0.12\\ \hline
         MINOS~\cite{Dorman:2009zz} & 1.19$(Q^2>0)$ &  World Average & 1.026$\pm$ 0.021~\cite{Bernard:2001rs} \\ 
   & 1.26$(Q^2>0.3GeV^2)$ & &\\ 
    &  &   & 1.014$\pm 0.014$~\cite{Bodek:2007ym} \\ 
   & & &\\ \hline
  \end{tabular}
\end{center}
\caption{Recent measurements of the axial dipole mass($M_A$).}
 \label{tab:axial_mass}
\end{table}
 
  The value of axial dipole mass $M_A$ used in Eq.~\ref{fa} has come recently in debate due to large deviations found in the experiments like
   MiniBooNE~\cite{AguilarArevalo:2010cx, AguilarArevalo:2010zc, AguilarArevalo:2013hm}, SciBooNE~\cite{Nakajima:2010fp}, K2K~\cite{Gran:2006jn}, etc. 
    from the world average value~\cite{Bernard:2001rs, Bodek:2007ym}. 
     Earlier measurements for $M_A$ were obtained using $\nu_l$ and $\bar\nu_l$ induced processes on deuterium 
     targets where the nuclear effects are expected to play negligible
     role and the average value is quoted as $M_A=1.026 \pm 0.021~GeV$ while a combined analysis 
     performed by Bodek et al.~\cite{Bodek:2007ym} using $\nu_{\mu}d$, $\overline\nu_{\mu}\text{H}$ and $\pi^\pm$ electroproduction 
    data have found $M_A=1.014 \pm 0.014~\text{GeV}$. Recent measurements at 
    NOMAD~\cite{Lyubushkin:2008pe} and MINER$\nu$A~\cite{Fiorentini:2013ezn} are close to the world average value. 
    On the other hand,  the experiments like 
    MiniBooNE~\cite{AguilarArevalo:2010cx, AguilarArevalo:2010zc, AguilarArevalo:2013hm},
  SciBooNE~\cite{Nakajima:2010fp}, K2K~\cite{Gran:2006jn}, etc. report a higher value of $M_A$. 
  These experiments were performed with different nuclear targets like $^{12}C$, $^{16}O$, $^{56}Fe$
    as well as some of them have used the
  same nuclear target, for example, NOMAD~\cite{Lyubushkin:2008pe}, MiniBooNE~\cite{AguilarArevalo:2010cx, AguilarArevalo:2010zc, AguilarArevalo:2013hm} 
  K2K~\cite{Gran:2006jn}, MINER$\nu$A~\cite{Fiorentini:2013ezn} have used carbon as nuclear target. 
  In Table-\ref{tab:axial_mass}, we tabulate the values of axial dipole mass obtained from
    analyses of some recent cross section measurements~\cite{Fields:2013zhk, Fiorentini:2013ezn,Nakajima:2010fp,Lyubushkin:2008pe,Gran:2006jn,AguilarArevalo:2010cx, AguilarArevalo:2010zc, AguilarArevalo:2013hm
 ,Bernard:2001rs,Dorman:2009zz} in the neutrino experiments in the few GeV energy range. 
    It is believed that if quasielastic like events which arise due to 2p-2h excitations, meson exchange currents
    and multinucleon correlations are taken into account then the recent experimental results can also be considered to 
    be consistent with a smaller value of $M_A$~\cite{Nieves:2011yp,Martini:2011wp,Martini:2012uc,Ankowski:2014yfa,Lalakulich:2012hs}. 
 However, it may be observed from Table-\ref{tab:axial_mass} that even with the same nuclear target 
  different values of $M_A$ have been obtained.
  
To study the explicit dependence of cross sections   
on axial dipole mass for neutrino/antineutrino scattering processes, 
 we have changed $M_A$ from the base value (taken as the world average value) and obtained 
the results for $\delta_{M_A}$ defined below in Eq. \ref{sdelta} and
 $\Delta_{M_A}$ defined in Eq. \ref{bdelta} by  taking the two different values of  
$M_A$ as $0.9~GeV$ and $1.2~GeV$. These results are obtained for the (anti)neutrino  induced processes on free nucleon as well as in the LFG with 
and without RPA effect for $^{40}Ar$ nuclear target. 

\begin{figure}
\includegraphics[height=10 cm, width=12.5 cm]{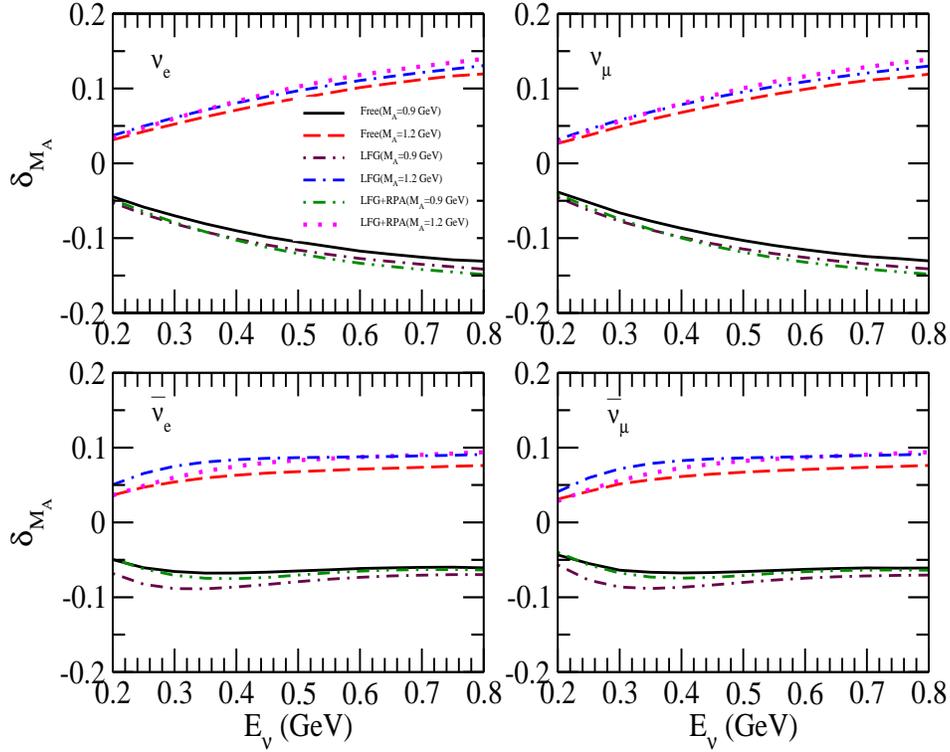}
  \caption{The dependence of cross section on $M_A$ obtained using Eq.~\ref{sdelta}. 
  The results are shown for $\nu_e(\bar\nu_e)$ and $\nu_\mu(\bar\nu_\mu)$ induced processes on free nucleon as well as 
  on $^{40}Ar$ target using LFG with and without RPA effect. Solid(dashed) line denotes results for the free nucleon case with $M_A=0.9~GeV$($1.2~GeV$), results obtained
  using LFG are shown by dashed dotted(double dashed dotted) with $M_A=0.9~GeV$($1.2~GeV$) and results 
  for LFG with RPA effect are shown by dashed double dotted(dotted) with $M_A=0.9~GeV$($1.2~GeV$).}
     \label{fig:partaldel_ma}
\end{figure}

 The dependence on axial dipole mass is shown in Fig.\ref{fig:partaldel_ma}, by defining $\delta_{M_A}$ as
 \begin{equation}\label{sdelta}
 \delta_{M_A}=\frac{\sigma_{\nu_l}(M_A^{modified}) - \sigma_{\nu_l}(M_A= 
WA)}{\sigma_{\nu_l}(M_A= WA)} ,\qquad \qquad \rm {\it WA= 1.026~GeV}
 \end{equation}
 where $l=e$ or $l=\mu$.
 We observe from Fig.\ref{fig:partaldel_ma} that for free nucleon when a modified value of  $M_A$ i.e. $M_A^{modified}= 0.9(1.2) ~GeV$ is used
 instead of world average
value of $1.026~GeV$ then a decrease(increase) of $5-15\%$ is obtained for $\nu_e / \nu_\mu$ reactions in the energy range of 
$0.2~GeV$ to $0.8~GeV$. In the case of $\bar\nu_e / \bar\nu_\mu$-nucleon reactions this decrease(increase) is about $5-10\%$ in
the same energy range. When nuclear medium effects are taken into account, for example, in the case of $^{40}Ar$ nucleus this
decrease(increase) remains almost same. Therefore, the uncertainty in the (anti)neutrino-nucleus cross sections is the same as in the case of free
(anti)neutrino-nucleon scattering processes. 

\begin{figure}
\includegraphics[height=8 cm, width=12.5 cm]{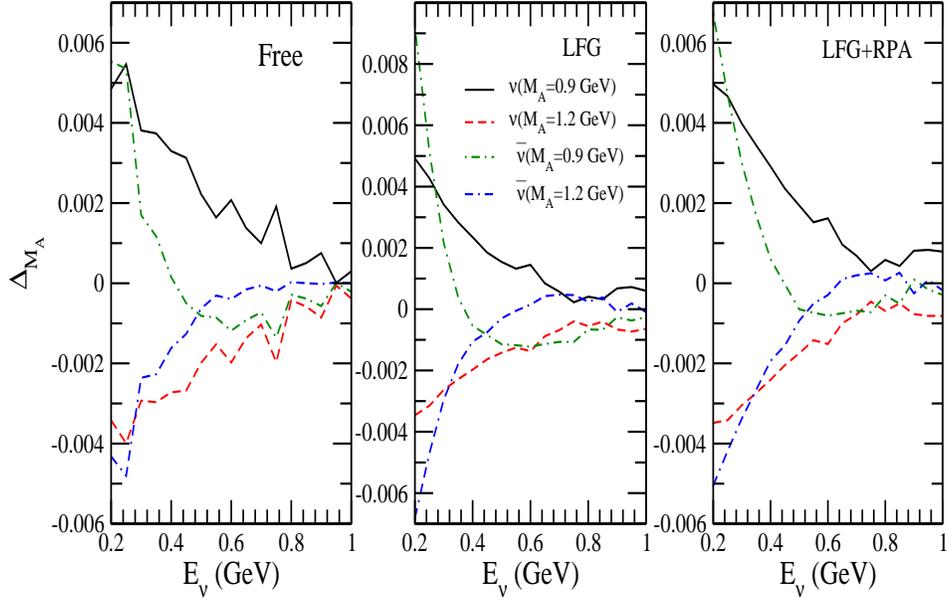}
  \caption{Effect of axial dipole mass on the cross section(from left to right): 
  on free nucleon; LFG, with and without RPA effect on $^{40}Ar$ target. Here different values of $M_A$ are 
  taken such as $0.9~GeV$ and $1.2~GeV$.
  The fractional difference(Eq.~\ref{bdelta}) has been obtained using the base value of $M_A$ taken as the world average value.
  Solid(dashed) line denotes results for the neutrino induced processes while for antineutrino the 
  results are shown by dashed dotted(double dashed dotted)  with $M_A=0.9~GeV$($1.2~GeV$).}
   \label{fig:del_ma}
\end{figure}

  We now study the sensitivity of the difference in electron and muon production cross sections due to the uncertainty in the choice 
 of $M_A$. For this we define
  \begin{eqnarray}\label{bdelta}
 \Delta_1{(E_\nu}) &=& \frac{{\sigma_{\nu_\mu}}(M_A^{modified}) - 
{\sigma_{\nu_e}}(M_A^{modified})}{{\sigma_{\nu_e}}(M_A^{modified})}, \nonumber\\
 \Delta_2{(E_\nu}) &=& \frac{{\sigma_{\nu_\mu}}(M_A = WA) - {\sigma_{\nu_e}}(M_A 
= WA)}{{\sigma_{\nu_e}}(M_A = WA)}, \nonumber\\
 \Delta_{M_A} &=& \Delta_1(E_\nu) - \Delta_2(E_\nu).
\end{eqnarray}
and show the numerical values for $\Delta_{M_A}$ for free nucleon and nuclei in Fig.~\ref{fig:del_ma}.
We observe that for free nucleon some sensitivity to the difference in the electron and muon production cross sections exists at low energies
$E_{\nu} < 0.4~GeV$ which does not exceed 1\%. In  nuclear target, when nuclear medium effects are taken into account this 
sensitivity remains almost the same as in the case of free neutrino-nucleon scattering processes. 
Similar effects are found for the case of antineutrino induced reactions.

From Figs.~\ref{fig:partaldel_ma} and \ref{fig:del_ma}, it may be observed that the cross section is very sensitive 
to the choice of axial dipole mass $M_A$. Therefore, while calculating the charged lepton production cross sections, 
the value of $M_A$ should be carefully taken.
 

\begin{figure}
 \includegraphics[height=7 cm, width=12.5 cm]{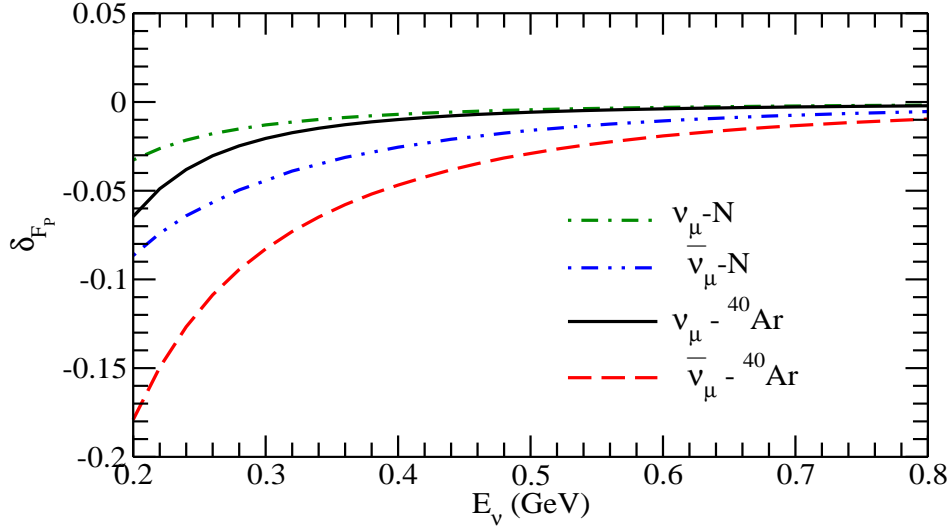}
 \caption{Results of the fractional change $\delta_{F_P}$ defined in 
 Eq.\ref{eq:deltaFpnew} as a function of (anti)neutrino energy. The results are shown for the $\nu_\mu$ induced 
 interaction cross section for the free nucleon case(dashed dotted line), as well as for the nucleons bound in $^{40}Ar$(solid line)
 nuclear target obtained by using LFG with RPA effect. The results corresponding to $\bar\nu_\mu$ induced
 CCQE process are shown by dashed double dotted line(free nucleon case) and dashed line($^{40}Ar$ target).}
 \label{fig:sdel_FpArnew}
\end{figure}
In the precision era of (anti)neutrino oscillation experiments looking for $\nu_\mu \leftrightarrow \nu_e$ or $\bar\nu_\mu \leftrightarrow \bar\nu_e$
 oscillation channels, it is important to understand the contribution from the vector part described in
 Eq.~\ref{eq:had_int} with $F_{1}^{V}(Q^2)$, $F_{2}^{V}(Q^2)$ form factors 
 and the axial-vector part with $F_{A}(Q^2)$, $F_{P}(Q^2)$ form factors of the first class current as well 
 as the contribution from the second class current with $F_{3}^{V}(Q^2)$ and  $F_{3}^{A}(Q^2)$ form factors. Furthermore, the contribution of the terms proportional 
 to lepton mass like the contribution from pseudoscalar form factor $F_P(Q^2)$ would be different for electron and muon channels.
 To observe their effect on lepton event rates, we have performed calculations using
  Eqs.~\ref{eq:deltaFp}, \ref{eq:delta2ndclass_F3V3} and \ref{eq:delta2ndclass_F3A3},
  in which we have studied the individual contribution of the terms in the hadronic current with and without $F_P(Q^2)$, $F_{3}^{V}(Q^2)$, $F_{3}^{A}(Q^2)$ terms.
\subsubsection{Pseudoscalar form factor}
To study the effect of pseudoscalar form factor $F_{P}(Q^2)$ on muon production cross sections, we define
\begin{equation}\label{eq:deltaFpnew}
   \delta_{F_P}{(E_\nu}) = \frac{{\sigma_{\nu_\mu}}(F_{P}\neq0) - {\sigma_{\nu_\mu}}(F_{P}=0)}{{\sigma_{\nu_\mu}}(F_{P}=0)},
  \end{equation}
and similar expression for antineutrino is used. For the numerical calculations expression of $F_P(Q^2)$ given in Eq.~\ref{fp} has been used. 
 The results are presented in Fig.\ref{fig:sdel_FpArnew}.
We find that $\delta_{F_P}$ is more sensitive in the case of $\bar\nu_\mu$ induced CCQE process than $\nu_\mu$ induced
process for the free nucleon case as well as for $^{40}Ar$ nuclear target. This sensitivity decreases with the increase in $\nu_\mu/\bar\nu_\mu$ energy
and almost vanishes beyond $0.6~GeV$. Using the three different expressions of the  pseudoscalar form factor $F_{P}(Q^2)$ given in 
Eqs. \ref{fp}, \ref{fp1} and \ref{fp2}, we have studied the behavior of $F_{P}(Q^2)$ vs $Q^2$ and found hardly any difference in the $Q^2$ dependence 
and therefore the total scattering cross section for $\nu_\mu$ and $\bar\nu_\mu$ induced processes are also unaffected by the choice of  $F_{P}(Q^2)$.

\begin{figure}
 \includegraphics[height=7 cm, width=12.5 cm]{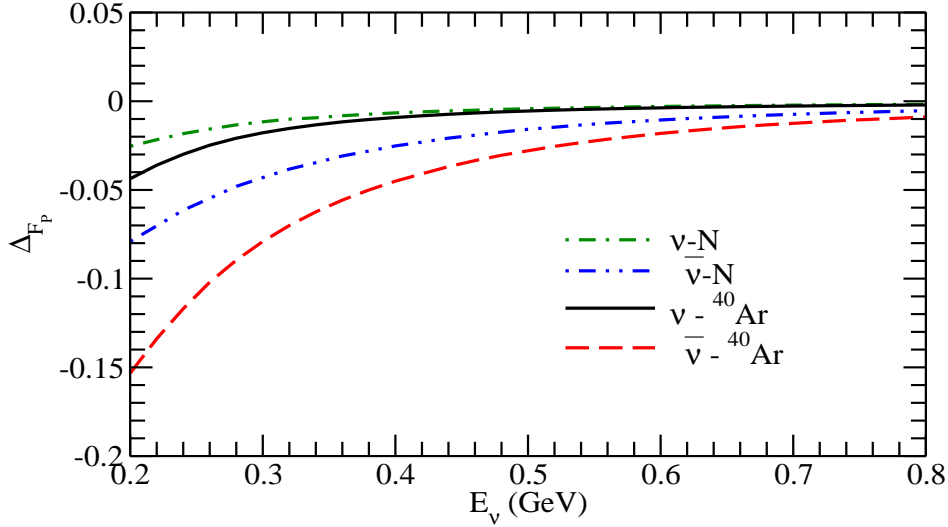}
 \caption{Results of the fractional change $\Delta_{F_P}$ defined in 
 Eq.\ref{eq:deltaFp} as a function of (anti)neutrino energy. The results are shown for the neutrino induced 
 interaction cross section for the free nucleon case(dashed dotted line), as well as for the nucleons bound in $^{40}Ar$(solid line)
 nuclear target obtained by using LFG with RPA effect. The results corresponding to antineutrino induced
 CCQE process are shown by dashed double dotted line(free nucleon case) and dashed line($^{40}Ar$ target).}
 \label{fig:del_FpAr}
\end{figure}

We also study the sensitivity of pseudoscalar form factor $F_{P}(Q^2)$ to find out the difference in the electron vs muon production cross sections
that are 
obtained using Eq.\ref{fp}. For this purpose we define
\begin{equation}\label{eq:deltaFp1}
   \Delta_1{(E_\nu}) = \frac{{\sigma_{\nu_\mu}}(F_{P}\neq0) - {\sigma_{\nu_e}}(F_{P}\neq0)}{{\sigma_{\nu_e}}(F_{P}\neq0)},
  \end{equation}
 \begin{equation}\label{eq:deltaFp2}
  \Delta_2{(E_\nu})= \frac{{\sigma_{\nu_\mu}}(F_{P}=0) - {\sigma_{\nu_e}}(F_{P}=0)}{{\sigma_{\nu_e}}(F_{P}=0)},
  \end{equation}
  \begin{equation}\label{eq:deltaFp}
  \Delta_{F_P} = \Delta_1(E_\nu) - \Delta_2(E_\nu).
  \end{equation}
 and the results for $\Delta_{F_P}$ are shown in Fig.~\ref{fig:del_FpAr}. Similar expressions are also used for antineutrino induced processes.

  We have calculated the fractional difference $\Delta_{F_P}$ as given in Eq.~\ref{eq:deltaFp} for free nucleon case 
as well as for nucleons bound in $^{40}Ar$ nuclear target using the LFG with RPA effect. 
We observe that the inclusion of pseudoscalar form factor decreases the  fractional change($\Delta_{F_P}$) by
about $3(8)\%$ at  $E_{\nu(\bar\nu)}$ $\sim 0.2GeV$ and becomes
smaller with the increase in energy. 
When the nuclear medium effects(LFG+RPA) are taken into account in the evaluation of cross sections in $^{40}Ar$ then this 
difference increases to $4(15)\%$ at the same energy for neutrino(antineutrino) induced processes. 

\subsection{Second class currents}
We have also studied the effect of second class current, due to which 
two additional form factors viz. $F_3^V(Q^2)$ and $F_3^A(Q^2)$ are introduced. 

\subsubsection{Second class vector current}
The contribution of the second class vector form factor $F_3^V(Q^2)$ to the cross section is always proportional to the mass of the lepton so it is quite small 
in the case of  $\nu_{e}(\bar \nu_e)$ as compared to $\nu_{\mu}(\bar \nu_\mu)$ reactions on free nucleons and nuclei. We first study the overall contribution
made by the second class vector form factor $F_3^V(Q^2)$ to the cross section in the case of muon neutrinos and define

\begin{figure}
 \includegraphics[height=7 cm, width=12.5 cm]{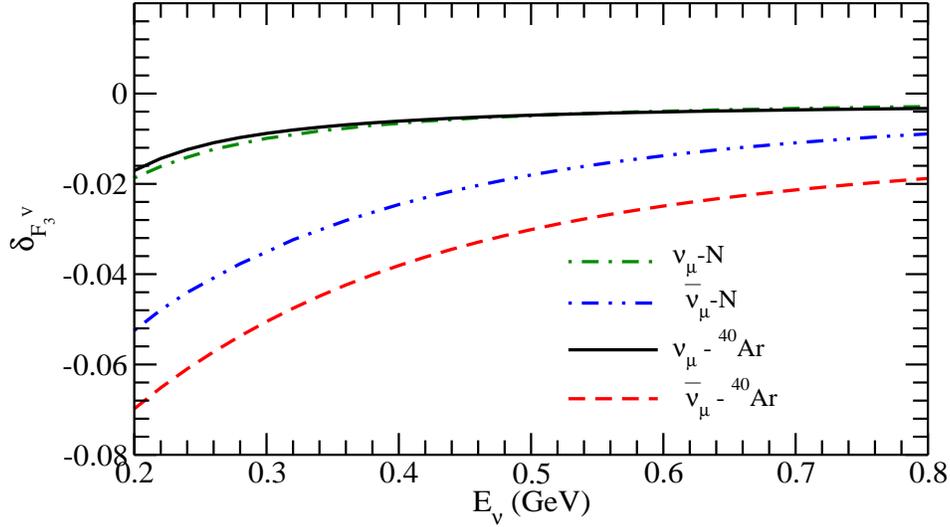}
 \caption{Results of the fractional change $\delta_{F_3^V}$ defined in 
 Eq.\ref{eq:deltaF3vnew} as a function of (anti)neutrino energy. The results are shown for $\nu_\mu$ induced 
 interaction cross section for the free nucleon case(dashed dotted line), as well as for the nucleons bound in $^{40}Ar$(solid line)
 nuclear target obtained by using LFG. The results corresponding to $\bar\nu_\mu$ induced
 CCQE process are shown by dashed double dotted line(free nucleon case) and dashed line($^{40}Ar$ target).}
 \label{fig:sdel_F3vArnew}
\end{figure}

\begin{equation}\label{eq:deltaF3vnew}
   \delta_{F_3^V}{(E_\nu}) = \frac{{\sigma_{\nu_\mu}}(F_{3}^{V}\neq0) - {\sigma_{\nu_\mu}}(F_{3}^{V}=0)}{{\sigma_{\nu_\mu}}(F_{3}^{V}=0)}.
  \end{equation}
Similar expression is used for antineutrino. For the numerical calculations we use Eq.~\ref{eq:f3v1} and the results are 
shown in Fig.\ref{fig:sdel_F3vArnew}. We find that the contribution of $F^V_3(Q^2)$ to the cross section 
is very small for $\nu_\mu$ scattering on free nucleons
 and nuclei. In the case of $\bar\nu_\mu$ scattering on nucleons at low energy, the contribution of $F^V_3(Q^2)$ 
 at $E_{\nu/\bar\nu}$ = 0.2 GeV is $5\%$ which
 increases to $7\%$ in $^{40}Ar$ when nuclear medium effects are taken into account.

\begin{figure}
 \includegraphics[height=8 cm, width=12.5 cm]{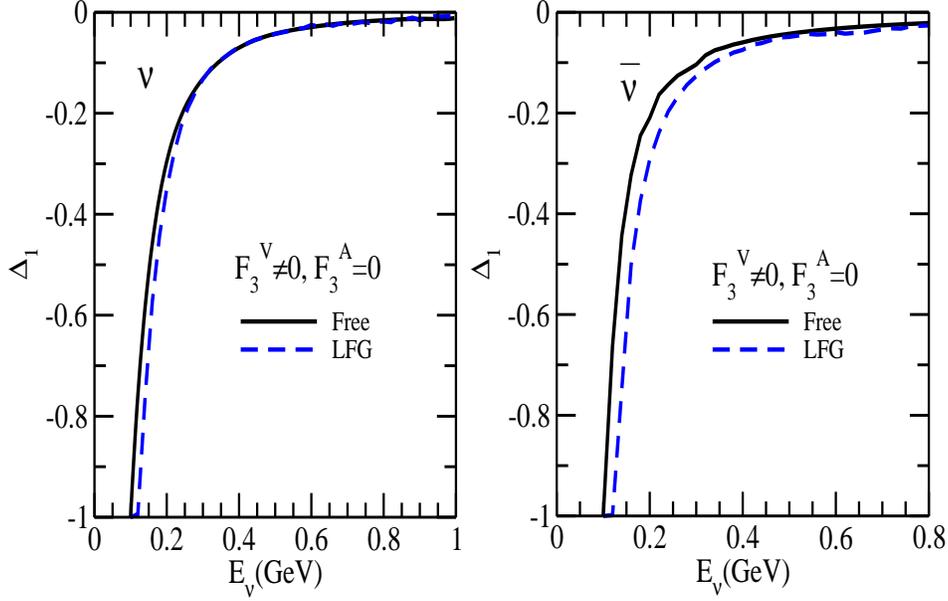}
 \caption{The difference of fractional changes $\Delta_{1}$ defined in 
Eqs.\ref{eq:delta2ndclass_F3V1}, for the free nucleon case(solid line), and in $^{40}Ar$ for neutrino(Left panel) and antineutrino(Right panel) induced processes 
using LFG(dashed line).}
 \label{fig:del_F3V2}
\end{figure}

We now study the sensitivity due to $F_3^V(Q^2)$ in the difference between the electron and muon
production cross sections for free nucleon and nuclei.
Sensitivity that arises in electron and muon production cross sections due to the presence of $F_3^V(Q^2)$ is studied by defining 
\begin{eqnarray}\label{eq:delta2ndclass_F3V1}
 \Delta_1{(E_\nu}) &=& \frac{{\sigma_{\nu_\mu}}(F_3^{V} \neq 0) - 
{\sigma_{\nu_e}}(F_3^{V} \neq 0)}{{\sigma_{\nu_e}}(F_3^{V} \neq 0)} \\
 \Delta_2{(E_\nu}) &=& \frac{{\sigma_{\nu_\mu}}(F_3^{V} = 0) - {\sigma_{\nu_e}}(F_3^{V} 
= 0)}{{\sigma_{\nu_e}}(F_3^{V} = 0)} \label{eq:delta2ndclass_F3V2}\\
 \Delta_{F_3^{V}} &=& \Delta_1(E_\nu) - \Delta_2(E_\nu).\label{eq:delta2ndclass_F3V3}
\end{eqnarray}
First, we present the results for $\Delta_1{(E_\nu})$ as a function of neutrino/antineutrino energies and the results are shown in Fig.~\ref{fig:del_F3V2}.
These results are presented for the free nucleon case as 
well as bound nucleons in $^{40}Ar$ and the cross sections are obtained using the expression of the hadronic current with second class vector current. We must point out 
that the contribution from the second class axial current is switched off. We find that $\Delta_1({E_\nu})$ is sensitive to the flavor of neutrinos especially at low 
energies ($E_{\nu/\bar\nu}~<~0.3GeV$) which is mainly due to threshold effect. 
When we perform calculations on nuclear targets like $^{40}Ar$ using LFG, the fractional difference $\Delta_1({E_\nu})$ changes
from free nucleon case.
However, it is not sensitive to the choice of nuclear model. 

\begin{figure}
 \includegraphics[height=8 cm, width=12.5 cm]{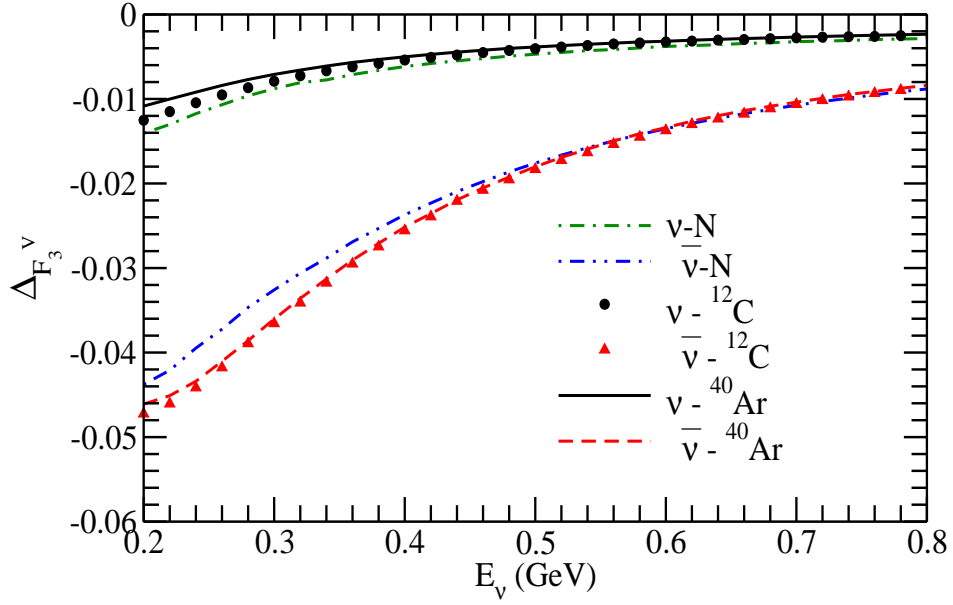}
 \caption{The difference of fractional changes $\Delta_{F_3^{V}}$ defined in 
Eq.\ref{eq:delta2ndclass_F3V3}, for the free nucleon case(neutrino results shown by dashed-dotted line  
and antineutrino results by dashed-double dotted line) as well as for $^{12}C$ 
(circle for neutrino and triangle up for antineutrino) and  
$^{40}Ar$ (solid line for neutrino and dashed line for antineutrino) nuclear targets obtained by using LFG.}
 \label{fig:del_F3V}
\end{figure}

In Fig.~\ref{fig:del_F3V}, we present the results for $\Delta_{F_3^V}$ using Eq.~\ref{eq:delta2ndclass_F3V3} and obtained it 
for the free nucleon case as well as for $^{12}C$ and $^{40}Ar$ nuclear targets.
 We find that the effects are energy dependent and are more pronounced at low energies. 
 From the figure it may be noticed that the fractional change is the same for 
both $^{12}C$ and $^{40}Ar$ nuclei.
For example, for the case of neutrino, at $E_\nu = 0.2~ GeV$, $\Delta_{F_3^V}$ is $\sim 1 \%$ for free nucleon as well as 
in $^{40}Ar$ evaluated using LFG and the difference
 becomes almost negligible beyond $E_{\nu}=0.5~GeV$.
Similarly, for the case of antineutrino, at $E_{\nu}  = 0.2~ GeV $, $\Delta_{F_3^V}$ is $\sim 4 \%$ for the free nucleon case 
 as well as in $^{40}Ar$ evaluated using LFG, which becomes $\sim 1 \%$  at $E_{\nu}  = 0.5~ GeV $. 
 
\begin{figure}
\includegraphics[height=8 cm, width=12.5 cm]{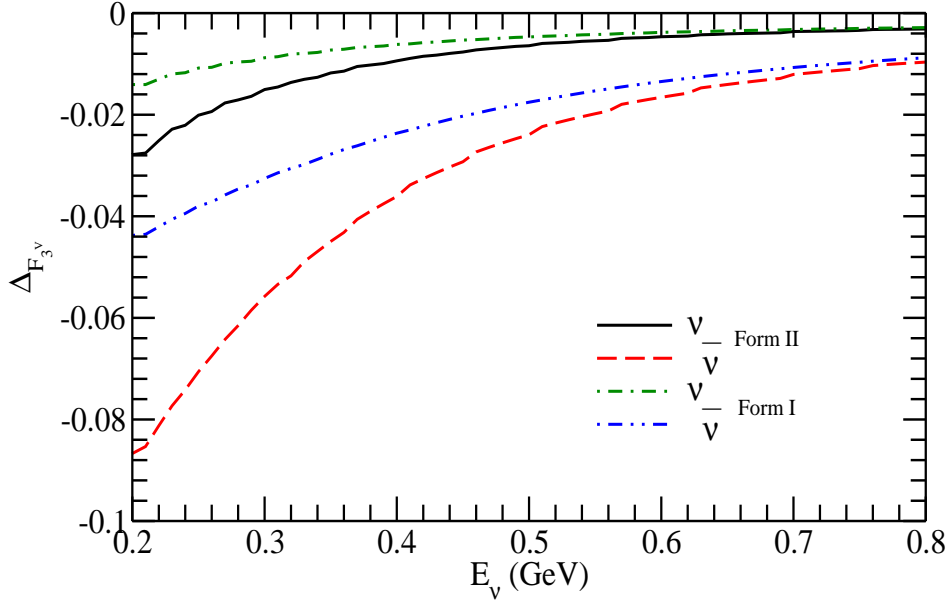}
 \caption{Variation of $\Delta_{F_3^V}$ as given in Eqs.~\ref{eq:delta2ndclass_F3V3} as a
  function of (anti)neutrino energies 
 are shown for different forms of $F_3^V(Q^2)$ used in second class currents. We have performed the calculation
 for (anti)neutrino induced scattering processes on free nucleon target. The results are obtained for Form I and Form II 
 using Eqs.~\ref{eq:f3v1}
 and \ref{eq:f3v2}, respectively for $M_A=1.026~GeV$. 
 The results obtained by using Form I for neutrino(antineutrino) are 
 shown by dashed-dotted(dashed-double dotted)line and with Form II are shown by solid(dashed)line, respectively.}
 \label{fig:del_2nd_forms}
\end{figure}

In general there is a large uncertainty associated with the
determination of $F_3^V(Q^2)$ and $F_3^A(Q^2)$ form factors. We have also studied uncertainty due to various parameterizations of form factor $F_3^V(Q^2)$.
Some of the alternative parameterizations of the form factor $F_3^V(Q^2)$ are given in Eqs.\ref{eq:f3v1},~\ref{eq:f3v2}, 
which have been used for numerical calculations. The results are  shown in Fig.\ref{fig:del_2nd_forms}.
   We find that for neutrino induced process on free nucleon target, 
the difference in the results for $\Delta F_{3}^{V}$ obtained by using two different forms of $F_3^V(Q^2)$
(Form I using Eq.~\ref{eq:f3v1}, Form II using Eq.~\ref{eq:f3v2}) is very small. 
For example, this difference is  $\sim 1\%$ at low energies($\sim 0.2~GeV$) which almost vanishes with the increase in energy. 
 In the case of antineutrino induced reaction on free nucleon, this difference is around $ 3\%$ at low energies which gradually vanishes with the increase in energy. 

\subsubsection{Second class axial current}
The axial vector form factor associated with second class current$(F_{3}^{A}(Q^2))$ also contributes to the cross section in
addition to the second class vector form factor$(F_{3}^{V}(Q^2))$.
To observe the effect of $F_{3}^{A}(Q^2)$ on $\nu_\mu$($\bar\nu_\mu$)
induced cross section we define fractional difference   
\begin{equation}\label{eq:deltaF3Anew}
   \delta_{F_3^A}{(E_\nu}) = \frac{{\sigma_{\nu_\mu}}(F_{3}^{A}\neq0) - {\sigma_{\nu_\mu}}(F_{3}^{A}=0)}{{\sigma_{\nu_\mu}}(F_{3}^{A}=0)},
  \end{equation}
  \begin{figure}
 \includegraphics[height=7 cm, width=12.5 cm]{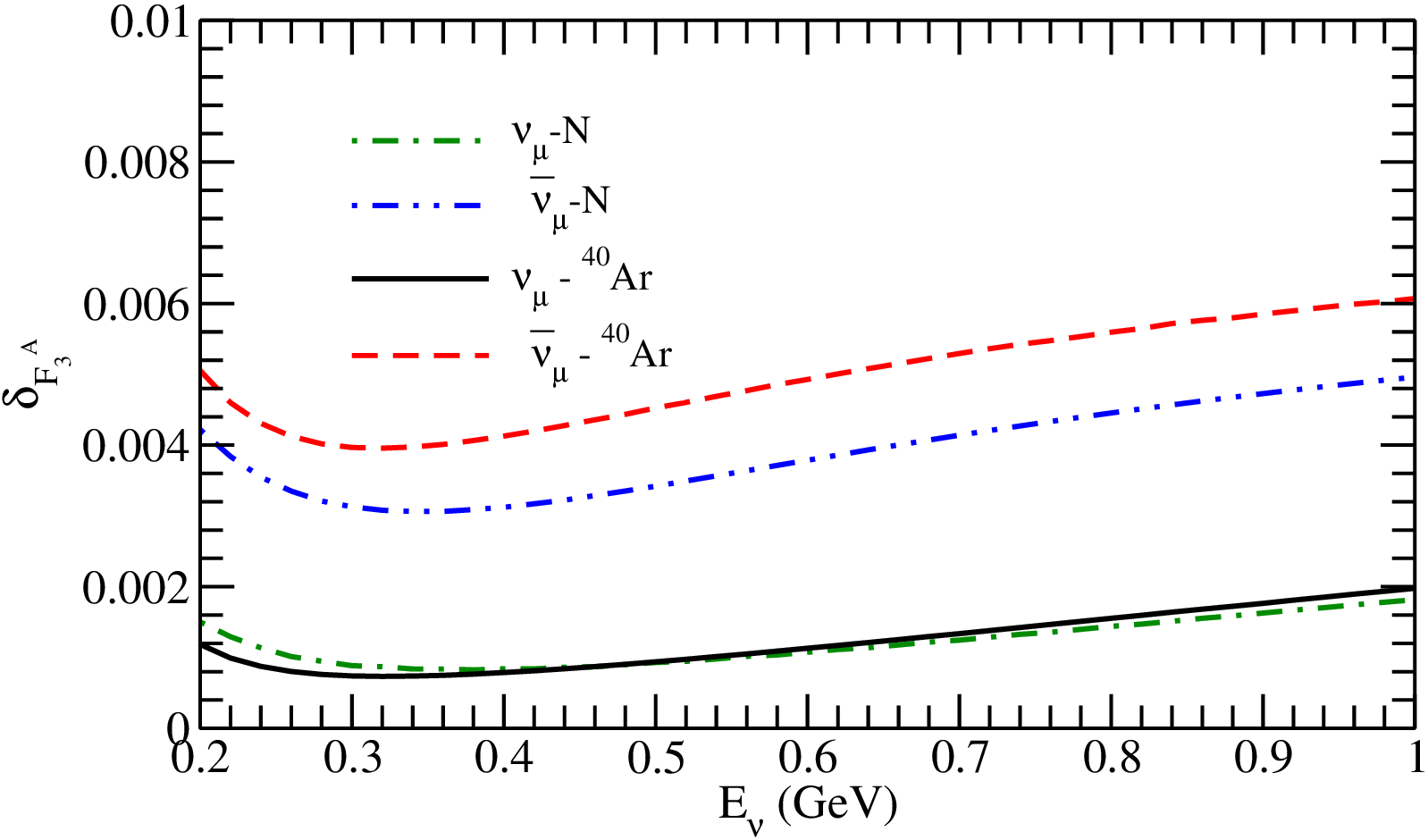}
 \caption{Results of the fractional change $\delta_{F_3^A}$ defined in 
 Eq.\ref{eq:deltaF3Anew} as a function of (anti)neutrino energy. The results are shown for the $\nu_\mu$ induced 
 interaction cross section for the free nucleon case(dashed dotted line), as well as for the nucleons bound in $^{40}Ar$(solid line)
 obtained by using LFG. The results corresponding to $\bar\nu_\mu$ induced
 CCQE process are shown by dashed double dotted line(free nucleon case) and dashed line($^{40}Ar$ target).}
 \label{fig:sdel_F3aArnew}
\end{figure}
and similar expression for antineutrino is used. We show the numerical results in Fig.\ref{fig:sdel_F3aArnew}.
These results are presented for the free nucleon case as well as for bound nucleons in $^{40}Ar$ and the cross sections are obtained using the expression of the 
hadronic current with second class axial vector current. This is to point out 
that the contribution from the second class vector current is switched off. We find that
$\delta_{F_3^A}$ is hardly sensitive  
 to the presence of $F_3^{A}(Q^2)$ in $\nu_\mu$ and  $\bar\nu_\mu$ scattering cross sections from free nucleon and nuclear targets.
We also study the sensitivity of the electron and muon production cross sections to $F_3^A(Q^2)$ for free nucleon and nuclei, 
by defining
\begin{eqnarray}\label{eq:delta2ndclass_F3A1}
 \Delta_1{(E_\nu}) &=& \frac{{\sigma_{\nu_\mu}}(F_3^{A} \neq 0) - 
{\sigma_{\nu_e}}(F_3^{A} \neq 0)}{{\sigma_{\nu_e}}(F_3^{A} \neq 0)}\label{eq:delta2ndclass_F3A2} \\
 \Delta_2{(E_\nu}) &=& \frac{{\sigma_{\nu_\mu}}(F_3^{A} = 0) - {\sigma_{\nu_e}}(F_3^{A} 
= 0)}{{\sigma_{\nu_e}}(F_3^{A} = 0)} \\
 \Delta_{F_3^{A}} &=& \Delta_1(E_\nu) - \Delta_2(E_\nu) \label{eq:delta2ndclass_F3A3}
\end{eqnarray}
and the numerical results are shown in Fig.\ref{fig:del_F3A1}, for $\Delta_1({E_\nu})$. We find the results to be similar in nature as found in the case 
of currents with second class vector form factor. In Fig.~\ref{fig:del_F3A}, we present the results for $\Delta_{F_3^{A}}$. In this case also we find the sensitivity
 to be smaller than observed in the case of $\Delta_{F_3^{V}}$.

\begin{figure}
 \includegraphics[height=8 cm, width=12.5 cm]{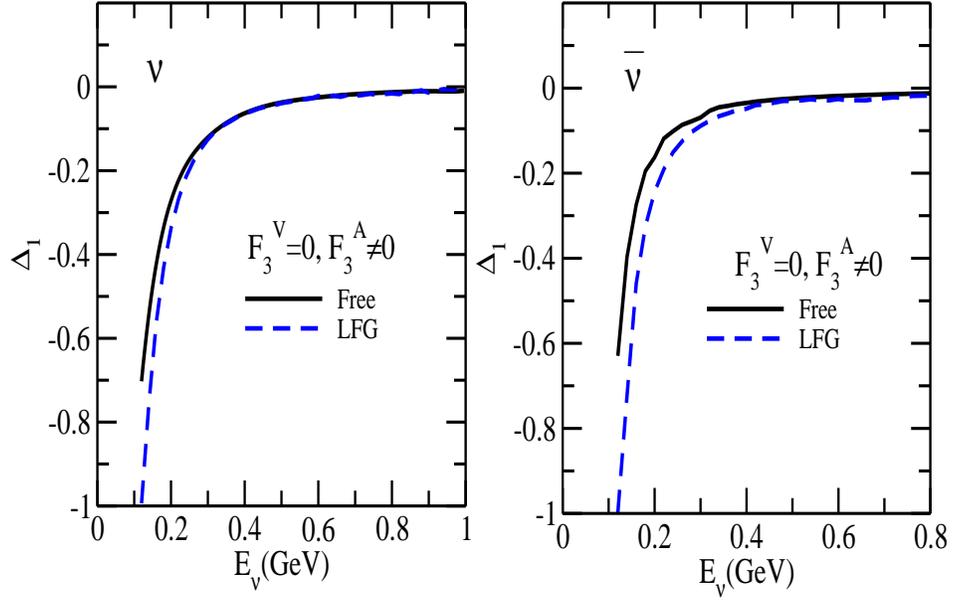}
 \caption{The difference of fractional changes $\Delta_{1}$ defined in 
Eqs.\ref{eq:delta2ndclass_F3A1}, for the free nucleon case(solid line), and in $^{40}Ar$ for neutrino(Left panel) and antineutrino(Right panel) induced processes 
using LFG(dashed line).}
 \label{fig:del_F3A1}
\end{figure}

 \begin{figure}
 \includegraphics[height=8 cm, width=12.5 cm]{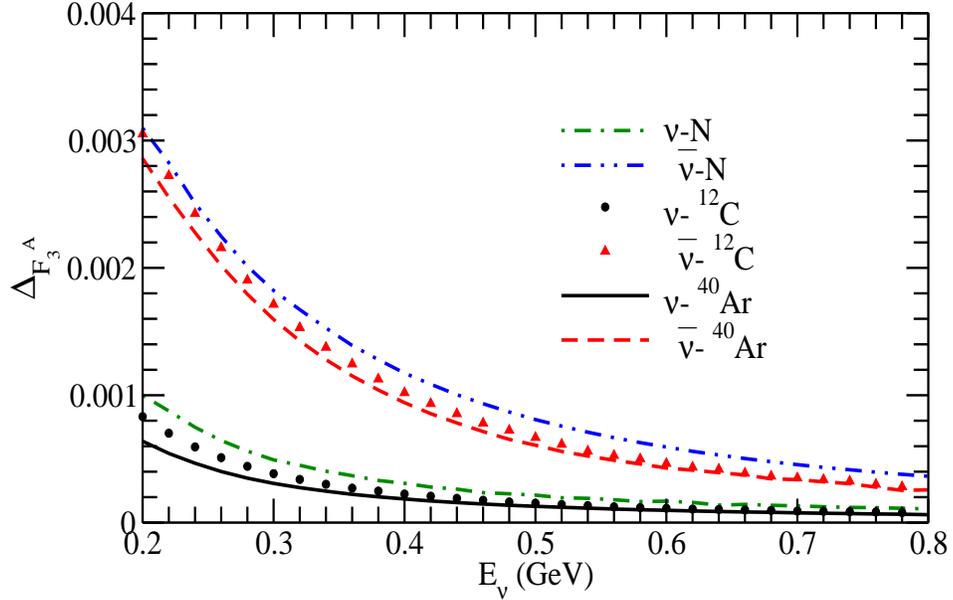}
 \caption{The difference of fractional changes $\Delta_{F_3^{A}}$ defined in 
Eq.\ref{eq:delta2ndclass_F3A3}, for the free nucleon case(neutrino results shown by dashed-dotted line  
and antineutrino results by dashed-double dotted line) as well as for $^{12}C$ (circle for neutrino and triangle up for antineutrino) and  
$^{40}Ar$ (solid line for neutrino and dashed line for antineutrino) obtained by using LFG.}
 \label{fig:del_F3A}
\end{figure}
When we compare our present results for the difference in the electron and muon production cross sections on
free nucleon target with the results of Day and McFarland~\cite{Day:2012gb}, we find that our results for the contribution of 
the $F_3^V(Q^2)$ in the case of antineutrino reactions and the results for the contribution of $F_3^A(Q^2)$ for neutrino
reactions agree qualitatively with their results. This is not so in the case of $F_3^V(Q^2)$ for neutrino reactions and 
$F_3^A(Q^2)$ for antineutrino reactions. This may be due the different expressions used for the contribution of the interference terms between first
 and second class currents. Our expressions agree with the general expressions given by Pais~\cite{Pais:1971er},
 Kuzmin et al.~\cite{Kuzmin:2007kr} but not with Eq.(3.18) of Llewellyn Smith~\cite{LlewellynSmith:1971zm} which has been used in Ref.~\cite{Day:2012gb}.
\begin{figure}
 \includegraphics[height=8 cm, width=12.5 cm]{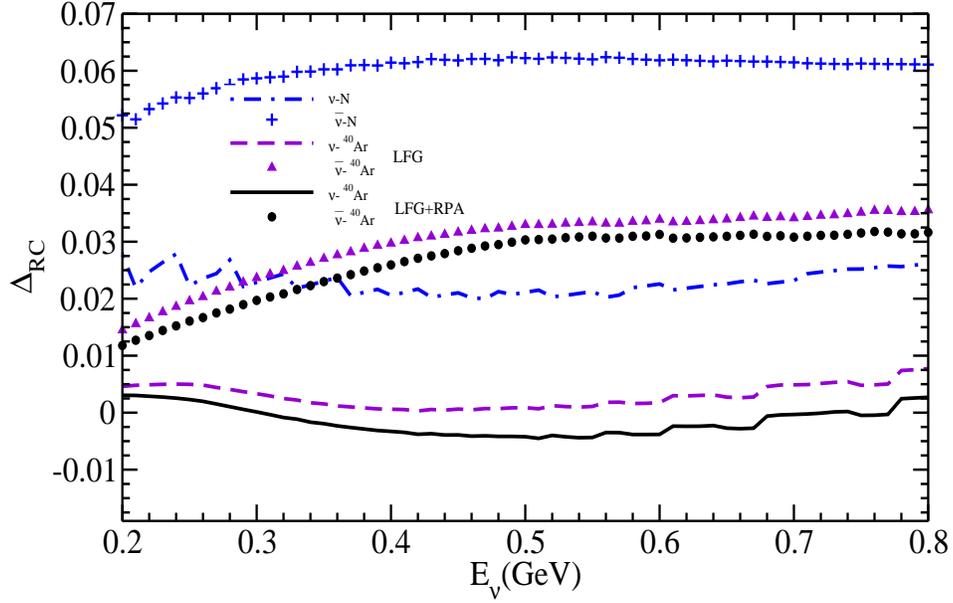}
 \caption{The effect of radiative corrections on fractional difference $\Delta_{RC}$ 
 defined in Eq. \ref{eq:delta_RC} for (anti)neutrino 
induced processes on free nucleon as well as on $^{40}Ar$ target using LFG with and without RPA effect. For the neutrino(antineutrino) induced processes on
 free nucleon the results are shown by dashed dotted line(plus), results for calculations using LFG for neutrino(antineutrino) are shown by dashed line
 (triangle up)  and results for calculations using LFG with RPA effect for neutrino(antineutrino) are shown by solid line(circle).}
  \label{fig:del_RC}
\end{figure}  
 \subsection{Radiative corrections}
 Radiative corrections are potential source of difference between electron and muon production cross sections in (anti)neutrino reactions due
 to their logarithmic dependence on the lepton mass through terms like log($\frac{E_l^\ast}{m_l}$), where $E_l^\ast$ is some energy scale in the reaction. 
 The radiative corrections in the charged current quasielastic neutrino-nucleon reactions relevant for the present oscillation experiments in the energy region of 
 few $GeV$ have been recently calculated by Bodek~\cite{Bodek:2007wb}, Day and McFarland~\cite{Day:2012gb} and Graczyck~\cite{Graczyk:2013fha}.
  Bodek~\cite{Bodek:2007wb} and Day and McFarland~\cite{Day:2012gb} make use of leading log approximation given by De Rujula et al.~\cite{De_Rujula:1979jj}
to calculate the contribution of soft  photon emission by the lepton leg bremsstrahlung diagram which gives major contribution 
 to the radiative corrections depending on the lepton mass $m_l$. 
 On the other hand, Graczyck~\cite{Graczyk:2013fha} includes the contribution of other diagrams like two boson exchange involving W and $\gamma$, 
 propagator correction in addition to the soft photon bremsstrahlung. We have used the results of De Rujula et al.~\cite{De_Rujula:1979jj} for the radiative corrections 
 in the neutrino nucleon scattering and study the effect of nuclear medium on these radiative corrections. 
 In the work of De Rujula et al.~\cite{De_Rujula:1979jj}, the modified cross section including radiative corrections is given by
\begin{eqnarray}\label{eqn:deRlll}
\frac{d\sigma}{dE_l d\Omega_l}&\approx&\frac{d\sigma_{free}}{dE_l d\Omega_l}
 + \frac{\alpha }{2\pi}\log\frac{4E^{*2}_l}{m_l^2}\int^1_0dz\frac{1+z^2}{1-z}
\left( \frac{1}{z}\frac{d\sigma_{free}}{d\hat{E}_l
  d\Omega_l}\left| _{_{\hat{E}_l=\frac{E_l}{z}}} \right. \theta(z- z_{min}) -\frac{d\sigma_{free}}{dE_l
  d\Omega_l}\right),~~~~~
\end{eqnarray}
where $\sigma_{free}$ is the (anti)neutrino induced cross section obtained without radiative effects, 
$E^*_l$ is the lepton energy in the center of mass of neutrino nucleon system and  $z_{min}$ is given by
\begin{eqnarray}
z_{min} &=& \frac{4 E_l^{\ast 2}}{2 M E_\nu}  \\ 
{\rm with} \quad E_l^{\ast} &= & \frac{s+m^2_l - M^2}{2 \sqrt s}
\end{eqnarray}
where $s=(k+p)^2$ is square of total energy in the center of mass system. $k$ and $p$ are the 
four momenta of incoming neutrino and target nucleon.

To show the effect of radiative corrections on the lepton event rates, we have obtained 
 total scattering cross sections for $\nu_e$ and $\nu_\mu$ induced reactions 
 on free and bound nucleon, with and without radiative corrections and define 
  \begin{eqnarray}\label{eq:delta_RC}
 \Delta_1{(E_\nu}) &=& \frac{{\sigma_{\nu_\mu}}(RC) - 
{\sigma_{\nu_e}}(RC)}{{\sigma_{\nu_e}}(RC)} \nonumber\\
 \Delta_2{(E_\nu}) &=& \frac{{\sigma_{\nu_\mu}}(NR) - 
{\sigma_{\nu_e}}(NR)}{{\sigma_{\nu_e}}(NR)} \nonumber\\
 \Delta_{RC} &=& \Delta_1(E_\nu) - \Delta_2(E_\nu)
\end{eqnarray}
where ${\sigma_{\nu_l}}(RC), \; (l=e, \mu)$ represents the cross sections obtained 
 by taking radiative corrections into account and ${\sigma_{\nu_l}}(NR), \; (l=e, \mu)$ represents
 the cross sections without radiative corrections.
 A similar definition has been used for antineutrino induced process. 
The results for $\Delta_{RC}$ are presented in Fig.~\ref{fig:del_RC}. 
These results are shown for the free nucleon target as well as for per interacting nucleon in $^{40}Ar$,
for which the total cross sections are calculated in LFG with and without RPA effect. 
We find that the difference in electron and muon production cross section $\Delta_{RC}$ due to radiative 
corrections is quite small in case of neutrino induced reactions as compared to antineutrino reactions. 
Furthermore, the effect of nuclear medium is to further decrease $\Delta_{RC}$  and the reduction is 
larger in case of antineutrino reactions as compared to neutrino reactions.
For example, for the neutrino induced process on free nucleon the effect is around 
$2 \%$ at $E_\nu = 0.5~ GeV$ while for 
antineutrino the effect is around $5 \%$ at $E_{\nu} = 0.5 ~GeV$.
When we performed the calculations using LFG with and without RPA effect, the effect 
becomes less than $1 \%$ in the case of neutrinos. 
For antineutrino induced process, using LFG without RPA effect, 
it is around $3 \%$, which becomes $\sim 2 \%$ when RPA effect is included.

\section{Summary and Conclusions}\label{sec:summary}
In the present study we observe the following:
\begin{enumerate}
 \item $\nu_\mu(\bar\nu_\mu)$ induced cross sections in free nucleon as well as nucleons bound in nuclear targets 
 are more suppressed due to threshold effects at low energies 
  than  $\nu_e(\bar\nu_e)$ induced reaction cross sections. Moreover, when cross sections are evaluated in nuclear targets there is a further reduction in the cross 
  sections due to nuclear medium effects. This reduction is energy dependent.
   For example, when calculations are performed for $^{12}C$ nucleus in LFG, the cross section for $\nu_e(\bar\nu_e)$ induced scattering 
   is reduced by $\sim 30(45)\%$ at $E_{\nu} = 0.3~GeV$  from free nucleon case while for $\nu_\mu(\bar\nu_\mu)$ induced scattering the reduction is 
   $\sim 32(46)\%$ at $E_{\nu} = 0.3~GeV$.     
 Inclusion of RPA correlation with LFG, further reduces the cross section from $\nu_e(\bar\nu_e)$ induced scattering by 
 $\sim 37(25)\%$ at 
 $E_{\nu} = 0.3~GeV$, which for  $\nu_\mu(\bar\nu_\mu)$ induced scattering is $\sim 40(27)\%$ at 
 $E_{\nu} = 0.3~GeV$. 
      This results in a larger difference in electron and muon production cross sections for the case of
     nuclear targets as compared to the free nucleon target.

 \item At low energies of $E_{\nu/\bar\nu} < 0.5~GeV$ there is appreciable nuclear model dependence on 
 (anti)neutrino-nucleus cross sections for both flavors of neutrino(antineutrino). The suppression
 due to nuclear medium effects is larger in the Local Fermi 
 Gas Model(LFG) as compared to the Fermi gas model of 
 Llewellyn Smith~\cite{LlewellynSmith:1971zm}. The suppression  in the Fermi gas models of 
 Smith and Moniz~\cite{Smith} and Gaisser and O'Connell~\cite{Gaisser:1986bv} are larger than LFG. 
 When RPA effect is included 
 in LFG, the suppression is largest.
 
   \item The suppression  due to nuclear medium effects is larger in the case of antineutrino as compared 
  to the cross sections obtained for neutrino induced processes. 
 
  \item For a given set of parameters which determine the form factors and other coupling constants 
  the percentage difference in electron and muon production cross sections 
  is more for nuclear targets than for the free nucleon target. This difference decreases with 
  neutrino/antineutrino energy. Also this difference increases
  with the increase in mass number. 
    
  \item The percentage difference in electron and muon production cross sections due to uncertainties in axial dipole 
  mass is more in the case of nuclear targets 
  as compared to free nucleon target. 
   The difference increases with the increase in mass number.
 
  \item  The fractional difference in the cross sections due to the presence of pseudoscalar form factor is more 
  in the case of $\bar\nu_\mu$ induced CCQE process than $\nu_\mu$ induced
process for the free nucleon case as well as in nuclear targets. Qualitatively at low neutrino energies there is a small difference in the results obtained for the free 
  nucleon target and nucleons bound in nucleus. This difference vanishes with the increase in energy. 
  In the case of antineutrino induced reaction the difference is slightly larger than found in the case of neutrino 
  and this difference does not vanish 
  with the increase in energy. The difference is almost independent of the choice of nuclear target.

 The contribution of pseudoscalar form factor in $\nu_\mu(\bar\nu_\mu)$ nucleon scattering is about $3(9)\%$
  at low energies of $E_{\nu}=0.2~GeV$ and becomes $\sim1(3)\%$ at $E_{\nu}=0.4~GeV$ in the
  case of free nucleon. When nuclear medium effects are taken into account this contribution increases.
    
     \item The inclusion of second class vector current results in an increase in the total scattering cross section if present 
     experimental limits of the second class form factor $F_3^V(Q^2)$ is used for $\nu_e(\bar\nu_e)$ as well as $\nu_\mu(\bar\nu_\mu)$
     induced processes. With the inclusion of $F_3^V(Q^2)$,  $\nu_\mu(\bar\nu_\mu)$ induced scattering cross section increases
     about $1(4)\%$ for free nucleon case and $1(8)\%$ when we include RPA with LFG at $E_{\nu}=0.4~GeV$. For the 
     $\nu_e(\bar\nu_e)$ induced processes this effect is smaller than in comparison to the $\nu_\mu(\bar\nu_\mu)$ induced processes.
     The difference in electron and muon 
      production cross sections increases at low energies for nuclear targets as compared to the free nucleon target. 
      This difference almost vanishes
      with the increase in neutrino energy. In the case of antineutrino induced CCQE process this difference 
      is slightly more as compared to the neutrino case. 
     However, this difference is almost independent of the choice of nuclear target.
     
     The effect of including second class axial vector form factor $F_3^A(Q^2)$  (consistent with present experimental limits) is
     qualitatively similar to the effect of including second class vector form factor but quantitatively quite small as compared to the effect of $F_3^V(Q^2)$
 summarized above.

   \item The effect of radiative corrections being proportional to $log(\frac{E_l^\ast}{m_l})$ affects the 
   $\nu_e(\bar\nu_e)$ scattering cross section  more than $\nu_\mu(\bar\nu_\mu)$ scattering cross sections when a corresponding charged lepton is in the final state.
   This gives 
   a difference in $\nu_e(\bar\nu_e)$ and $\nu_\mu(\bar\nu_\mu)$ scattering cross sections which is almost independent
   with energy in the case of neutrino induced process while it 
   increases slightly with energy in the case of antineutrino
   induced process both for the free nucleon as well as the bound nucleons. 
   We find this difference to decrease in the presence of nuclear medium effects and this 
   decrease is more in the case of antineutrino as compared to the neutrino reactions. 
   \end{enumerate}
\subsection{Acknowledgments}
MSA is thankful to Prof. Kevin S. McFarland and some other participants of 
NuInt14 for showing their interest in this work. MSA is also grateful to 
the Department of Science and Technology (DST), Government of India,
for providing financial assistance under Grant No.SR/S2/HEP-18/2012. FZ acknowledges Maulana Azad National
 Fellowship.

\appendix

\section{Hadronic tensor $J^{\mu\nu}$}
\subsection{Relativistic expression for the hadronic tensor $J^{\mu\nu}$}
We have followed the prescription of Nieves et al.~\cite{Nieves:2004wx} while taking RPA correlations into account. 
The difference is that we have also included the contribution coming due to the second class currents. The following is 
the expression for the hadronic tensor $J^{\mu\nu}$:
\begin{eqnarray}
 J^{\mu\nu}&=& 4\left[(F_1^V)^2 (p^\mu q^\nu +q^\mu p^\nu
 + 2 p^\mu p^\nu+ q^2 g^{\mu\nu}/2)-(F_2^V)^2\left\{
 \frac{- q^2g^{\mu\nu}}{2}+ \left(1+\frac{q^2}{4M^2}\right)
 \frac{q^\mu q^\nu}{2}\right.\right.\nonumber\\
 &+&\left.\left.\frac{q^2}{4M^2}\left(p^\mu q^\nu +q^\mu p^\nu +
 2 p^\mu p^\nu\right)\right\} + 
 (F_A)^2 \left\{2M^{2}g^{\mu\nu}\left(\frac{q^2}{4 M^2} -1\right) + 
  q^\mu p^\nu+p^\mu q^\nu+2 p^\mu p^\nu\right\}\right. \nonumber\\
  &-& \left.(F_P)^2 \frac{q^2 q^\mu q^\nu}{2M^2} + 2(F_3^V)^2 
  \left(1-\frac{q^2}{4 M^2}\right) q^\mu q^\nu -  (F_3^A)^2 \frac{q^2}{2M^2}
  \left(2p^\mu+q^\mu\right)
  \left(2p^\nu+q^\nu\right)\right.\nonumber \\
  &+& \left.F_1^V F_2^V\left(q^2 g^{\mu\nu}-q^\mu 
  q^\nu\right)-2 i(F_1^V F_A + F_2^V F_A)\epsilon^{\mu\nu\alpha\beta}
  p_\alpha q_\beta -2 F_A F_P q^\mu q^\nu \right.\nonumber\\
  &-&  \left.F_3^A F_P\frac{q^2}{M^2}(p^\mu q^\nu +q^\mu p^\nu +  
  q^\mu q^\nu)- 2 F_3^A F_A \left(p^\mu q^\nu +q^\mu p^\nu + 
  q^\mu q^\nu\right) \right.\nonumber\\
  &+& \left.2 F_1^V F_3^V \left(p^\mu q^\nu +q^\mu p^\nu +q^\mu q^\nu \right)+
  F_2^V F_3^V \frac{q^2}{2M^2} ( p^\mu q^\nu +q^\mu p^\nu + 
  q^\mu q^\nu)\right]
\end{eqnarray}

\subsection{Component form of $J^{\mu\nu}$ incorporating RPA in the lowest order}
Here the three momentum transfer ${\vec q}$ is taken along z-axis and RPA is applied in the leading terms. The different 
components of the hadronic tensors 
$J^{\mu\nu}$ with RPA effect are 
\begin{eqnarray}
J^{00}_{\rm RPA} &=& 4\left[ 2(F_1^V)^2 
\left(\underline{\bf C_N}\,E^2(\vec{p}\,)
+\frac{q^2}{4}+q^0E(\vec{p}\,)\right)  - \frac{q^2}{2 }
(F_2^V)^2 \left ( \frac{ (q^0)^2}{q^2} + \frac{\vec{p}^{\,2}+q^0E(\vec{p})+ (q^0)^2/4}{M^2}\right )\right.
\nonumber \\ 
&+& \left. 2 (F_A)^2 \left ( q^0E(\vec{p})+ \frac{q^2}{4}+\vec{p}^{\,2}\right )
-\underline{\bf C_L}\, (F_P)^2 \frac{q^2 q_0^2}{2 M^2}
+2(F_3^V)^2  (q^0)^2 \left(1-\frac{q^2}{4 M^2}\right) \right. \nonumber\\
&-&\left. (F_3^A)^2 \frac{q^2}{2}\left(\frac{q^0 + 2 E(\vec{p})}{M}\right)^2
-\underline{\bf C_N} F_1^V F_2^V \vec{q}^{\,2}
+ 2F_1^V F_3^V ( (q^0)^2 + 2 E(\vec{p})q^0)\nonumber \right. \\
&+& \left. F_2^V F_3^V \frac{q^2}{M^2}\left(E(\vec{p}) q^0 + \frac{(q^0)^2}{2}\right)
- 2F_P F_A (q^0)^2 - 4 F_A F_3^A \left(E(\vec{p})q^0 + (q^0)^2\right)\right. \nonumber \\
&-& \left. 2 F_P F_3^A \frac{q^2}{M^2}\left(E(\vec{p})q^0 + (q^0)^2\right)\right]
\end{eqnarray}
\begin{eqnarray}
J^{0z}_{\rm RPA} &=& 4\left[  (F_1^V)^2\left
(\underline{\bf C_N} E(\vec{p}\,)(2p_z+|\vec{q}\,|)+
q^0p_z\right ) - \frac{q^2}{4}(F_2^V)^2    \left (
\frac{E(\vec{p}\,)}{M}\frac{2p_z+|\vec{q}\,|}{M}
 + 2\frac{q^0 |\vec{q}\,|}{q^2}
+ \frac{q^0 (2p_z+|\vec{q}\,|)}{2M^2} \right)\right. \nonumber\\
&+& \left. (F_A)^2 \left (
\underline{\bf C_L} E(\vec{p}\,)(2p_z+|\vec{q}\,|)+ 
 q^0p_z\right) -\underline{\bf C_L}\, (F_P)^2 q^0 |\vec{q}| \frac{q^2}{2M^2}
 + 2(F_3^V)^2 q^0|\vec{q}| 
 \left(1-\frac{q^2}{4M^2}\right) \right. \nonumber\\
 &-& \left. (F_3^A)^2 \frac{q^2}{2 M^2}(|\vec{q}|
+2 p_z)(q^0+ 2 E(\vec{p}))-F_1^V  F_2^V 
q^0 |\vec{q}\,|+2 F_1^V F_3^V(E(\vec{p})|\vec{q}|+
q^0 p_z+q^0|\vec{q}| )\right. \nonumber\\
&+&\left. F_2^V F_3^V \frac{q^2}{2M^2}(E(\vec{p})|\vec{q}|+
q^0 p_z+q^0|\vec{q}| )
- 2 F_P F_A q^0|\vec{q}| -2 F_A F_3^A(E(\vec{p})|\vec{q}|+ 
q^0 p_z+q^0|\vec{q}| ) \right. \nonumber\\
&-&\left. F_P F_3^A\frac{q^2}{M^2}(E(\vec{p})|\vec{q}|
+ q^0 p_z+q^0|\vec{q}| )\right]\\
J^{zz}_{\rm RPA} &=& 4\left[ 2 (F_1^V)^2\left( 
 p_z^2+|\vec{q}\,|p_z-\frac{q^2}{4}\right)
-\frac{q^2}{2}( F_2^V)^2 \left ( \left
(\frac{2p_z+|\vec{q}\,|}{2M}\right)^2 + \frac{(q^0)^2}{q^2} \right )\right.\nonumber\\ 
&+&\left. 
2(F_A)^2 M^2 \left ( \underline{\bf C_L} +
\frac{p_z^2+|\vec{q}\,|p_z-q^2/4}{M^2}\right) -\underline{\bf
C_L}\,(F_P)^2\frac{q^2 \vec{q}^2}{2 M^2}+ 2(F_3^V)^2 \vec{q}^2 \left(1-\frac{q^2}{4M^2}\right)\right. \nonumber \\
&-&\left. (F_3^A)^2 \frac{q^2}{2 M^2}\left(2 p_z+|\vec{q}|\right)^2 -(q^0)^2 F_1^V F_2^V 
+ 2 F_1^V F_3^V (2 p_z |\vec{q}|+\vec{q}^2)
+ F_2^V F_3^V \frac{q^2}{2M^2}(2p_z |\vec{q}| + \vec{q}^2) \right. \nonumber \\
&-& \left. F_P F_3^A \frac{q^2}{M^2}(2p_z |\vec{q}| + \vec{q}^2)-2 F_A F_P\vec{q}^2
-2 F_A F_3^A(2p_z |\vec{q}| + \vec{q}^2)\right] \\
J^{xx}_{\rm RPA} &=& 4 \left[2 (F_1^V)^2 \left
(p_x^2-\frac{q^2}{4} \right ) -\frac{q^2}{2}( F_2^V)^2 \left
(\underline{\bf C_T}+\frac{p_x^2}{M^2} \right )
+ 2F_A^2 M^2 \left (
\underline{\bf C_T}  +
\frac{p_x^2-q^2/4}{M^2}\right)\right. \nonumber \\ 
&-&\left. 2 (F_3^A)^2 \frac{q^2 p_x^2}{M^2} -\underline{\bf C_T} q^2 F_1^V F_2^V \right]\\
J^{xy}_{\rm RPA} &=& - 8{\rm i} F_A M^2 (F_1^V+ F_2^V) 
\left ( \underline{\bf C_T} |\vec{q}\,|E(\vec{p}\,)-q^0 p_z\right )
\end{eqnarray}

\begin{equation}
{\bf C_N} = \frac{1}{|1-c_0f^\prime(\rho)U_N(q,k_F)|^2}, \quad
{\bf C_T} = \frac{1}{|1-U(q,k_F)V_t(q)|^2}, \quad
{\bf C_L} = \frac{1}{|1-U(q,k_F)V_l(q)|^2} \label{eq:coeffs}\nonumber
\end{equation}
where $V_l$ and $V_t$ are the longitudinal and transverse part of the nucleon-nucleon potential calculated with $\pi$ and $\rho$ exchanges and are given by
\begin{eqnarray}
V_l(q) &=& \frac{f^2}{m^2_\pi}\left
\{\left(\frac{\Lambda_\pi^2-m_\pi^2}{\Lambda_\pi^2-q^2 }\right)^2
\frac{\vec{q}{\,^2}}{q^2-m_\pi^2} + g^\prime\right \}, \quad
\frac{f^2}{4\pi}=0.08,~~\Lambda_\pi=1.2~{GeV},~~m_\pi=0.14~{GeV} \nonumber \\
V_t(q) &=& \frac{f^2}{m^2_\pi}\left
\{   C_\rho \left (\frac{\Lambda_\rho^2-m_\rho^2}{\Lambda_\rho^2-q^2 }\right)^2
\frac{\vec{q}{\,^2}}{q^2-m_\rho^2} + g^\prime \right \},
~C_\rho=2,\Lambda_\rho=2.5~{GeV}, m_\rho=0.77~{GeV} \label{eq:st2}
\end{eqnarray}
$g^\prime$ is the Landau-Migdal parameter taken to be 0.7 which has been used quite successfully to explain
many electromagnetic and weak processes in nuclei. $U(q,k_F)=U_N(q,k_F)+U_\Delta(q,k_F)$ is the Lindhard function for the particle-hole excitation and 
 $U_\Delta(q,k_F)$ is the Lindhard function for the delta-hole excitation. The details are given in Ref.\cite{Nieves:2004wx,Oset1}.

\end{document}